\documentclass[journal]{IEEEtran}

\usepackage{colortbl}
\usepackage{graphicx}
\usepackage{stfloats}
 \usepackage{subfigure} 
\usepackage{cite}

\ifCLASSINFOpdf

\else

\fi

\usepackage[cmex10]{amsmath}
\usepackage{amssymb}
\usepackage{tikz}
\usepackage{enumerate}

\begin{document}

	\title{On Secure Degrees of Freedom of the MIMO Interference Channel with Local Output Feedback}
	
	\author{Tong~Zhang,~\IEEEmembership{Member,~IEEE,}
		Yinfei~Xu,~\IEEEmembership{Member,~IEEE,}
	 Shuai~Wang,~\IEEEmembership{Member,~IEEE,} \\
	 Miaowen~Wen,~\IEEEmembership{Senior Member,~IEEE,}
	 and~Rui~Wang,~\IEEEmembership{Member,~IEEE}

		\thanks{T.~Zhang, S.~Wang, and R. Wang are with the Department of Electrical and Electronic Engineering, Southern University of Science and Technology, Shenzhen, China (email: bennyzhangtong@yahoo.com, wangs3@sustech.edu.cn, and wang.r@sustech.edu.cn).}
		\thanks{Y.~Xu is with the School of Information Science and Engineering, Southeast University, Nanjing, 210096, China (email: yinfeixu@seu.edu.cn).}
		\thanks{M.~Wen is with the School of Electronic and Information Engineering, South China University of Technology, Guangzhou 510641, China (e-mail: eemwwen@scut.edu.cn).}
		}

	\maketitle
	
\begin{abstract}
This paper studies the problem of sum-secure degrees of freedom (SDoF)  of the $(M,M,N,N)$ multiple-input multiple-output (MIMO) interference channel with local output feedback, so as to build an information-theoretic foundation and provide practical transmission schemes for 6G-enabled vehicles-to-vehicles (V2V). For this problem, we propose two novel transmission schemes, i.e., the interference decoding scheme and the interference alignment scheme, and thus establish a sum-SDoF lower bound. In particular, to optimize the phase duration, we analyze the security and decoding constraints and formulate a linear-fractional optimization problem. Furthermore, we show that the derived sum-SDoF lower bound is the sum-SDoF for $M \le N/2$, $N=M$, and $2N \le M$ antenna configurations, and reveal that for a fixed $N$, the optimal $M$ to maximize the sum-SDoF is not less than $2N$. Through simulations, we examine the secure sum-rate performance of proposed transmission schemes, and reveal that using local output feedback can lead to a higher secure sum-rate than that by using delayed channel state information at the transmitter. 
\end{abstract}
	

	\section{Introduction}
	
 \IEEEPARstart{T}{he} sixth-generation (6G) mobile networks are envisioned to provide high data rate (1Tbps), support high-mobility (1000kmph), and ensure physical-layer security  \cite{201,202,203}. To satisfy the unprecedented requirements of high data rate, high-mobility, and physical-layer security, the 6G-enabled Internet of Vehicles (IoV) has to face the following challenges:

\begin{itemize}	\item \textit{Challenge of Massive Multiple-input Multiple-output (MIMO)}: Massive MIMO enables the multiplexing of a large amount of data streams in the same time/frequency, which dramatically increases the data rate \cite{201,202,203,400}. However, the analysis of multiplexing gain is challenging, as it is subject to the number of antennas, network topology, and channel state information (CSI) \cite{00,70,121}. Since the multiplexing gain is also known as degrees-of-freedom (DoF), the DoF characterization for the 6G-enabled IoV is a fundamental research problem.  
	
	\item \textit{Challenge of Fast Fading Channel}: 	The 6G-enabled IoV is required to support the communications with mobility up to 1000kmph \cite{201,202,203}. A consequence of such high-mobility communications is to make the wireless channel fast fading. Hence, the feedback of CSI is easy to be delayed. Under delayed feedback, the transmission scheme is different from the celebrated transmission schemes with current CSI. To date, the transmission scheme design with delayed feedback is an open research problem. 
	
	\item \textit{Challenge of Employing Physical-layer Security}:  As privacy and security becoming increasingly important for mobile networks, the physical-layer security techniques plan to employ in 6G-enabled IoV \cite{201,202,203}. However, due to high mobility, the traditional physical-layer security techniques in the static scenario may not be effective in 6G-enabled IoV. Therefore, the physical-layer security techniques for 6G-enabled IoV remain to be investigated.

\end{itemize}

 The 6G-enabled vehicles-to-vehicles (V2V) is an important application of 6G-enabled IoV, where high-mobility vehicles transmit data to other high-mobility vehicles with large data rate and security guarantee. Likewise, the 6G-enabled V2V is with the challenges of massive MIMO, fast fading channel, and employing physical-layer security, as well. This motivates us to investigate the secure degrees-of-freedom (SDoF) characterization for the MIMO interference channel with local output feedback. Thereafter, we are able to establish an information-theoretic foundation and provide practical transmission schemes for 6G-enabled V2V.

It is well known that the SDoF is defined as the first-order approximation of secure channel capacity in the high signal-to-noise-ratio (SNR) regime. Mathematically, the secure channel capacity can be expressed by $\text{SDoF}\log_2\text{SNR} + o(\log_2(\text{SNR}))$, where $\lim_{\text{SNR} \rightarrow {\cal{1}}} o(\log_2(\text{SNR}))/\log_2(\text{SNR}) = 0$. Hence, the SDoF characterizes the fundamental limits of multiplexing for secure communications in a wireless network, where we can configure the number of data streams in a wireless network according to the value of SDoF as an upper limit.

The $(M,M,N,N)$ MIMO interference channel has two multiple-antenna transmitters and two multiple-antenna receivers, where all transmitters have $M$ antennas and all receivers have $N$ antennas, and the transmitter $i=1,2$ has a message for the receiver $i$. As an important type of delayed feedback, the local output feedback refers to the feedback of the outputs at the receiver $i=1,2$ to the transmitter $i$, which consists of delayed CSI and data symbols.

In this paper, we aim at exploring the sum-SDoF of the $(M,M,N,N)$ MIMO interference channel with local output feedback, and providing practical transmissions schemes. The main contributions of this paper are summarized as follows:
\begin{enumerate}
\item The $N/2 < M \le N$ antenna configurations imply that the receiver can immediately decode the $M$-dimensional desired signals and $(N-M)$-dimensional interfering signals. Motivated by the above observation, we propose a novel interference decoding scheme. In contrast to all the existing secure transmission schemes with delayed feedback \cite{20,21,22,23,24,25,26,27,28,29,402,30}, the novelty of our transmission scheme lies in the simultaneous data transmission enabled by interference decoding. This transmission design needs the security and decoding analysis, which involves determining the rank of related matrices with non-trivial structure. Built upon the security and decoding analysis, a linear-fractional optimization problem is established to find the optimal phase duration, which maximizes the sum-SDoF lower bound achieved by our scheme.
		
\item The $N < M \le 2N$ antenna configurations imply that the receiver cannot immediately decode the desired and interfering signals, and the re-transmission of interference will not create new interference but provide useful equations for decoding. Motivated by the above observation, we customize an interference alignment scheme. Unlike the separate artificial noise (AN) transmission in \cite{29} for delayed channel state information at the transmitter (CSIT), our transmission scheme adopts a joint AN transmission in the interference channel enabled by output feedback. Furthermore, we analyze the security and decoding constraints of the proposed transmission scheme. To maximize the sum-SDoF lower bound achieved by our scheme, a linear-fractional optimization problem is established to find the optimal phase duration.  
		
\item We evaluate and compare the performance of the results. Through the existing sum-SDoF upper bound, we show that the derived sum-SDoF lower bound is the sum-SDoF for $M \le N/2$, $N=M$, and $2N \le M$ antenna configurations. Moreover, the derived sum-SDoF lower bound is not less than the state-of-the-art sum-SDoF lower bound with delayed CSIT in \cite{29} for all antenna configurations. Through simulations, we examine the secure sum-rate performance of proposed interference decoding and interference alignment schemes. The simulation results show that compared with that of \cite{29}, our scheme can have a higher secure sum-rate.		 
\end{enumerate}
	
\textit{Notations}: The scalar, vector and matrix are denoted by $h,\textbf{h}$, and $\textbf{H}$, respectively. The identity matrix of dimensions $m$ is denoted by $\textbf{I}_m$. The Euclidean norm is denoted by $||\cdot||$. The conjugate-transpose is denoted by $(\cdot)^H$. The determinant 
  is denoted by $|\cdot|$. The rank of matrix \textbf{Y} is denoted by $\text{rk}\{\textbf{Y}\}$. The block-diagonal matrix with blocks $\textbf{X}$ and $\textbf{Y}$ is denoted by
 \begin{equation}
 	\text{bd}\{\textbf{X}, \textbf{Y}\} =
 	\begin{bmatrix}
 		\textbf{X} & \textbf{0} \\
 		\textbf{0} & \textbf{Y}
 	\end{bmatrix}.
 \end{equation}
 
\begin{figure}[h]
		\centering
		\includegraphics[width=1.75in]{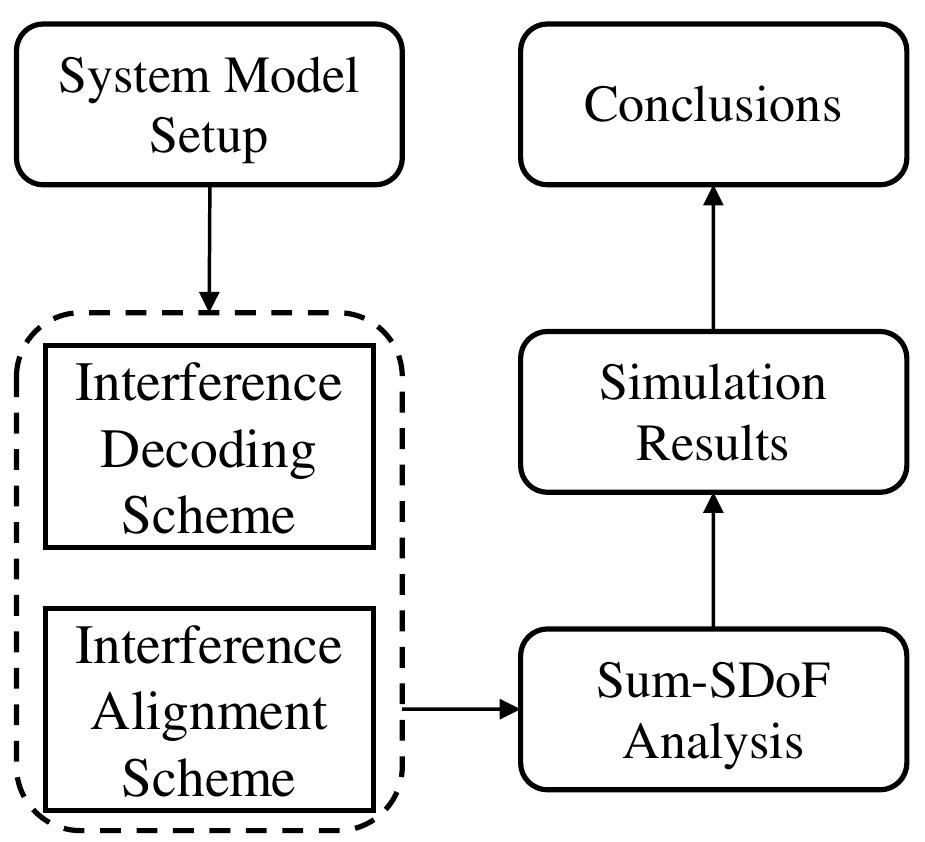}
		\caption{ Illustration of the research procedure of this paper.}
		\label{S0}
\end{figure}

\textit{Organizations}: The related work with emphasis on research gap is provided in Section-II. The system model is stated in Section-III. We design the transmission schemes for the $(M,M,N,N)$ MIMO interference channel with local output feedback in Section-IV. Then, we are able to prove the Theorem 1 and Corollary 1 to characterize the sum-SDoF of the $(M,M,N,N)$ MIMO interference channel with local output feedback  in Section-V. We examine the practical performance of the proposed transmission schemes via simulations in Section-VI. The conclusions are drawn in Section-VII. We illustrate the research procedure of this paper in Fig. \ref{S0}.
	
\section{Related Work}

The study of SDoF with delayed feedback was initiated from the scenario of broadcast channel \cite{20,21,22,23,24,25}. In \cite{20}, the SDoF region of the two-user MIMO broadcast channel with delayed CSIT was characterized, where the transmitter and receivers have arbitrary antennas. For the two-user broadcast channel with delayed CSIT, when there is an external eavesdropper, the SDoF inner and outer regions were derived in \cite{21}. Thereafter, in \cite{22}, the sum-SDoF of $K$-user multiple-input single-output (MISO) broadcast channel with delayed CSIT was obtained. The synergy of current CSIT, delayed CSIT, and no CSIT for broadcast channel was investigated from SDoF perspective in \cite{23,24}. With alternating current CSIT and delayed CSIT, the SDoF region of two-user MISO broadcast channel with an external eavesdropper was derived in \cite{23}. In \cite{24}, with alternating current CSIT, delayed CSIT, and no CSIT, the SDoF region of two-user MISO broadcast channel was characterized. The study of \cite{25} derived the sum-SDoF of the multiple-input multiple-output multiple-eavesdropper wiretap channel (MIMOME), where the transmitter wants to communicate with a legitimate receiver with delayed CSIT in the presence of many eavesdroppers with no CSIT.

Aside from the broadcast channel, the SDoF characterization was shifted to the X channel and interference channel, which have attracted many interests \cite{26,27,28,29,402,30}. For the single-input single-output (SISO) interference channel with delayed CSIT, the research can be found in \cite{26,27,28}. For the $K$-user SISO interference channel with delayed CSIT, a sum-SDoF lower bound was obtained in \cite{26,27}, where the receiver beamforming technique was utilized to handle multi-source of interference when the receiver has more antennas than that of transmitter. For the $2 \times 2 \times 2$ SISO interference channel with delayed CSIT, the sum-SDoF was characterized in \cite{28}.  For the MIMO interference channel, a sum-SDoF lower bound of the $(M,M,N,N)$ MIMO interference channel with delayed CSIT was proposed in \cite{29}, when each transmitter has $M$ antennas and each receiver has $N$ antennas. As for the X channel, the message setting is different from that of the interference channel. In particular, the transmitter of X channel has messages for two receivers, in which the interference is likely to be aligned at receivers, rather than decoded. For the $(M,M,N,N)$ MIMO X channel with delayed CSIT, a sum-SDoF lower bound was obtained in \cite{402}.
The research of \cite{30} showed that, for the $(M,M,N,N)$ MIMO X channel with output feedback, the same sum-SDoF as many as that in the two-user MIMO broadcast channel with delayed CSIT and output feedback can be achieved \footnote{With delayed CSIT, the transmitter in the MIMO broadcast channel can re-construct the output signals at the receivers. Hence, for the MIMO broadcast channel, the sum-SDoF with delayed CSIT is equal to the sum-SDoF with delayed CSIT and output feedback, where the former one is derived in \cite{20}.}. On the other hand, output feedback enjoys a much lower feedback overhead (i.e., the number of complex coefficients needed for feedback)  than that of delayed CSIT. To be specific, the feedback overhead of delayed CSIT is of an order of $MN$, while the feedback overhead of output feedback is of an order of $N$.

\begin{figure*}[t]
 		\centering
 		\includegraphics[width=6in]{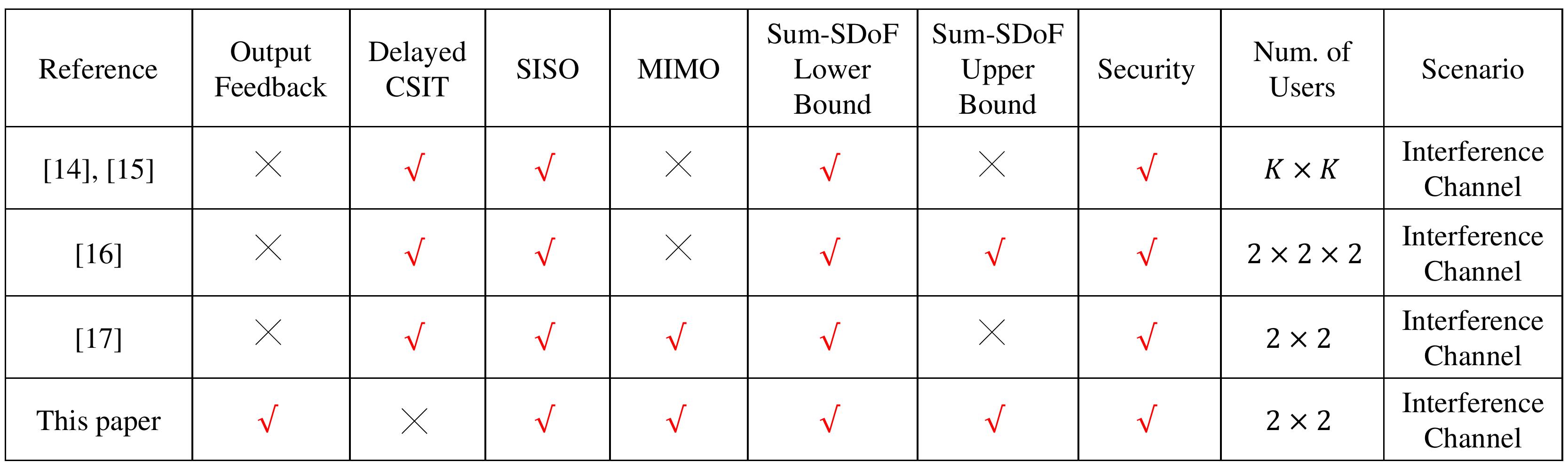}
 		\caption{ Comparison of the problem considered in this paper with related work in the interference channel.}
 		\label{S}
 	\end{figure*}

However, for the MIMO interference channel with output feedback, to the best of each author's knowledge, the SDoF characterization is still unexplored in the open literature. Motivated by this research gap, in this paper, we focus on the problem of SDoF characterization of the $(M,M,N,N)$ MIMO interference channel with local output feedback.
In Fig. \ref{S}, we compare the considered problem with related work in the interference channel.
 
\section{System Model and Problem Statement}

In this section, we first introduce the considered scenario of 6G-enabled V2V. Then, we set up the model of the $(M,M,N,N)$ MIMO interference channel with confidential messages and local output feedback. Next, we define the SDoF tuple. Finally, we state the considered sum-SDoF problem. For the ease of explanation, the important acronyms in this paper are summarized in Table I. 

\begin{table}[t]
	\centering
	\caption{Summary of Important Acronyms}
	\begin{tabular}{l|l}
		\hline 
		\rowcolor{gray!10}	  \textbf{Acronym} & \textbf{Full Name}   \\ \hline 
		DoF & Degrees of Freedom \\  \hline
		\rowcolor{gray!10} 		  SDoF & Secure Degrees of Freedom \\  \hline
		IoV & Internet of Vehicles  \\  \hline
		\rowcolor{gray!10}		 V2V & Vehicles-to-Vehicles  \\  \hline 
		SNR & Signal-to-Noise-Ratio \\ \hline
		\rowcolor{gray!10}	 	MISO & Multiple-Input Single-Output \\ \hline
		MIMO &  Multiple-Input Multiple-Output \\ \hline
		\rowcolor{gray!10}	 		CSI & Channel State Information \\ \hline
		CSIT & Channel State Information at the Transmitter \\ \hline 
			\rowcolor{gray!10}	  TS & Time Slot \\ \hline
	\end{tabular}
	\label{Tab2}
\end{table}

\subsection{The Considered Scenario of 6G-enabled V2V}

	\begin{figure}[b]
	\centering
	\includegraphics[width=3in]{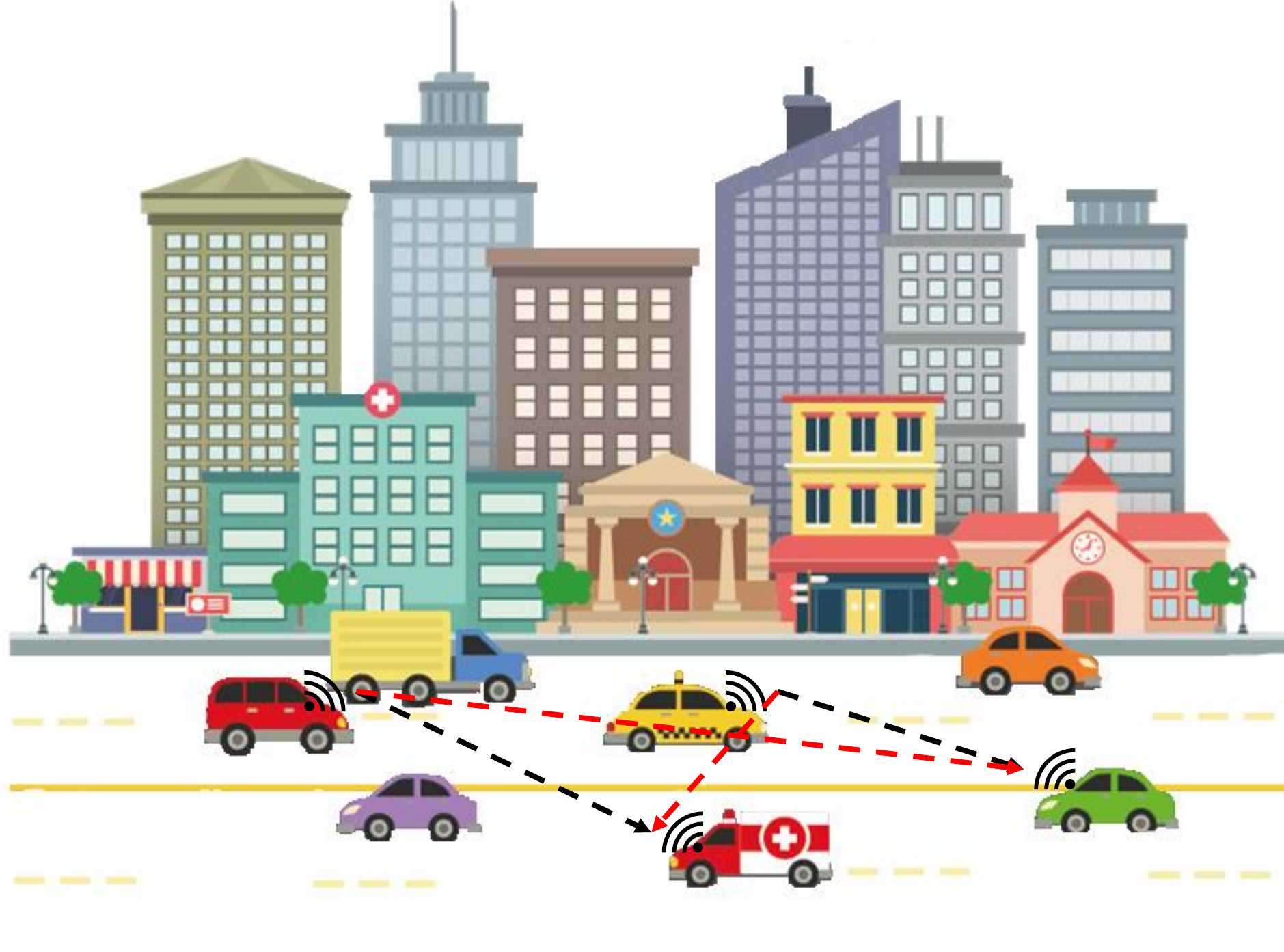}
	\caption{Considered scenario of 6G-enabled V2V. The desired signal link  and the interference signal link are colored in black and red, respectively.} \label{F1}
\end{figure}

We consider a 6G-enabled V2V communication scenario, depicted in Fig. \ref{F1}, where two vehicles simultaneously communicate with the other two vehicles. Due to the broadcast nature of the wireless medium, the signal transmission for the desired vehicle may incur harmful interference and potential information leakage to the other vehicle. In particular, each vehicle has a confidential message for its paired vehicle, which can be a personal video, a high-definition private image, a set of operation statuses, etc. Meanwhile, the wireless channel is fast fading, due to high-mobility. Accordingly, there are following three important assumptions behind this scenario:
\begin{enumerate}
	\item  Due to the buildings and trees alongside the road and many vehicles on the road, it is a rich-scatter environment for communications \cite{220,221,222,223}; 
	\item  All vehicles are equipped with multiple antennas, because massive MIMO is an effective way to meet high data rate requirement of 6G \cite{230,231,232}; and
	\item  Due to the high-mobility of vehicles, the wireless channel is fast fading, which leads to the delayed feedback from the receivers to the transmitters \cite{240,241,242}.
\end{enumerate}

In the following, it will be seen that such 6G-enabled V2V may fall into the MIMO interference channel with confidential messages and local output feedback.

\subsection{The $(M,M,N,N)$ MIMO Interference Channel with Confidential Messages and Local Output Feedback}
	
	 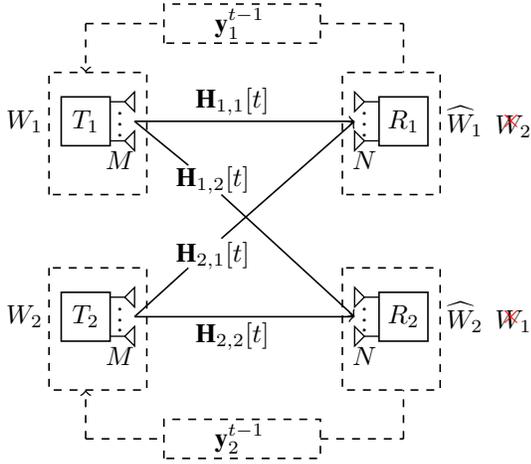
\begin{figure}[t]
		\begin{center}
			\begin{tikzpicture}[scale = 0.65]
				\node at (-1.5,0) {$T_2$};
				\node at (-2.75,0) {$W_2$};		
				\node at (-1.5,4) {$T_1$};
				\node at (-2.75,4) {$W_1$};		
				\node at (5,0) {$R_{2}$};
				\node at (6.25,0) {$\widehat{W}_{2}$}; 	
				\node at (7.25,-0.1) {$W_{1}$}; 		
				\node at (5,4) {$R_{1}$};
				\node at (6.25,4) {$\widehat{W}_{1}$};
				\node at (7.25,3.9) {$W_{2}$};
				\node at (7.2,4) {$\textcolor{red}{\times}$};
				\node at (7.2,0) {$\textcolor{red}{\times}$};

				\node at (1.5,-0.4) {$\textbf{H}_{2,2}[t]$};
				\node at (1.5,4.4) {$\textbf{H}_{1,1}[t]$};
				
				\node at (1.1,2.8) {$\textbf{H}_{1,2}[t]$};
				\node at (1.1,1.25) {$\textbf{H}_{2,1}[t]$};

				\draw[semithick,->] (-0.5,0) -- (4,0);
				\draw[semithick,->] (-0.5,4) -- (4,4);
				
				\draw[semithick] (-0.5,0) -- (0.6,1);
				\draw[semithick,->] (1.275,1.6)-- (4,4);
				
				\draw[semithick] (-0.5,4) --  (0.6,3.1);
				\draw[semithick,->] (1.275,2.5) --  (4,0);

				\draw [semithick] (-2,-0.5) rectangle (-1,0.5);
				\draw [semithick] (-2,3.5) rectangle (-1,4.5);
				
				\draw [semithick] (4.5,-0.5) rectangle (5.5,0.5);
				\draw [semithick] (4.5,3.5) rectangle (5.5,4.5);
				
				\draw [line width=0.5pt] (-1,4.4) -- (-0.7,4.4) -- (-0.5,4.6) -- (-0.5,4.2) -- (-0.7,4.4);
				\node at (-0.8,4.1) {$\vdots$};
				\draw [line width=0.5pt]  (-1,3.6) -- (-0.7,3.6) -- (-0.5,3.8) -- (-0.5,3.4) -- (-0.7,3.6);
				\node at (-0.8,3.2) {$M$};
				
				\draw [line width=0.5pt] (-1,0.4) -- (-0.7,0.4) -- (-0.5,0.6) -- (-0.5,0.2) -- (-0.7,0.4);
				\node at (-0.8,0.1) {$\vdots$};
				\draw [line width=0.5pt]  (-1,-0.4) -- (-0.7,-0.4) -- (-0.5,-0.2) -- (-0.5,-0.6) -- (-0.7,-0.4);
				\node at (-0.8,-0.8) {$M$};

				\draw [line width=0.5pt] (4.5,4.4) -- (4.2,4.4) -- (4,4.6) -- (4,4.2) -- (4.2,4.4);
				\node at (4.2,4.1) {$\vdots$};
				\draw [line width=0.5pt]  (4.5,3.6) -- (4.2,3.6) -- (4,3.8) -- (4,3.4) -- (4.2,3.6);
				\node at (4.2,3.2) {$N$};

				\draw [line width=0.5pt] (4.5,0.4) -- (4.2,0.4) -- (4,0.6) -- (4,0.2) -- (4.2,0.4);
				\node at (4.2,0.1) {$\vdots$};
				\draw [line width=0.5pt]  (4.5,-0.4) -- (4.2,-0.4) -- (4,-0.2) -- (4,-0.6) -- (4.2,-0.4);
				\node at (4.2,-0.8) {$N$};
				
				\draw [semithick,dashed] (-2.25,2.5) rectangle (-0.25,5);
				\draw [semithick,dashed] (3.75,2.5) rectangle (5.75,5);
				
				\draw[semithick,dashed,->] (-1.5,6) -- (-1.5,5);
				\draw[semithick,dashed] (-1.5,6) -- (0.1,6);
				
				\draw [semithick,dashed] (0.1,5.6) rectangle (3.3,6.4);
				\draw[semithick,dashed] (3.3,6) -- (5,6);
				\draw[semithick,dashed] (5,6) -- (5,5);
				
				\node at (1.65,6){$\textbf{y}_1^{t-1}$};

				\draw [semithick,dashed] (-2.25,1) rectangle (-0.25,-1.5);
				\draw [semithick,dashed] (3.75,1) rectangle (5.75,-1.5);
				\draw[semithick,dashed,->] (-1.5,-2.5) -- (-1.5,-1.5);
				\draw[semithick,dashed] (-1.5,-2.5) -- (0.1,-2.5);	
				
				\draw [semithick,dashed] (0.1,-2.9) rectangle (3.3,-2.1); 	
				\draw[semithick,dashed] (3.3,-2.5) -- (5,-2.5);
				\draw[semithick,dashed] (5,-2.5) -- (5,-1.5);
				\node at (1.65,-2.5){$\textbf{y}_2^{t-1}$};	
			\end{tikzpicture}
			\caption{$(M,M,N,N)$ MIMO interference channel with confidential messages and  local output feedback. }\label{P1}
		\end{center}
	\end{figure}
	
	We consider a $(M,M,N,N)$ MIMO interference channel has two transmitters and two receivers, i.e., transmitters 1, 2, and receivers 1, 2, where each transmitter has $M$ antennas and each receiver has $N$ antennas. The transmitter $1$ has a confidential message $W_1$ for receiver $1$, and the transmitter $2$ has a confidential message $W_2$ for receiver $2$. Receiver $i=1,2$ returns output signals to the transmitter $i$, i.e., local output feedback. The considered $(M,M,N,N)$ MIMO interference channel with confidential messages and local output feedback is depicted in Fig. \ref{P1}.

	Mathematically, the complex input signal at the transmitter $i=1,2$ and time slot (TS) $t$ is denoted by $\textbf{x}_i[t]$. The complex output signal at receiver $j=1,2$ and TS $t$ is denoted by $\textbf{y}_j[t]$. The input-output relationships are given by
	\begin{subequations}
		\begin{eqnarray}
		&& \textbf{y}_1[t] = \textbf{H}_{1,1}[t]\textbf{x}_1[t] + \textbf{H}_{2,1}[t]\textbf{x}_2[t] + \textbf{z}_1[t], \\
		&& \textbf{y}_2[t] =  \textbf{H}_{1,2}[t]\textbf{x}_1[t] + \textbf{H}_{2,2}[t]\textbf{x}_2[t] + \textbf{z}_2[t],
		\end{eqnarray}
	\end{subequations}
	where the CSI matrix from the transmitter $i=1,2$ to the receiver $j=1,2$ at TS $t$ is denoted by $\textbf{H}_{i,j}[t] \in \mathbb{C}^{N \times M}$, and the additive white Gaussian noise (AWGN) vector at the receiver $j$ and TS $t$ is denoted by $\textbf{z}_j[t]$. Due to the rich scatter environment, we are able to assume that $\textbf{H}_{i,j}[t]$ is full rank. Also, due to the fast fading wireless channel, we are able to assume that $\textbf{H}_{i,j}[t_1]$ is independent of $\textbf{H}_{i,j}[t_2]$ for $t_1 \ne t_2$.

\subsection{SDoF Tuple}
	
A secure code $(2^{nR_1(\text{SNR})},2^{nR_2(\text{SNR})},n)$ with secure  achievable rates $R_1(\text{SNR})$ and $R_2(\text{SNR})$ is defined as follows: The communication process takes $n$ channel uses. The confidential message at transmitter $i=1,2$ is denoted by $W_i=[1:2^{nR_i(\text{SNR})}]$.  A stochastic encoder $f_i(\cdot)$ at the transmitter $i$ encodes confidential message $W_i$ and local output $\textbf{y}_i^{t-1} = [\textbf{y}_i[1],\cdots,\textbf{y}_i[t-1]]$ to a codeword $\textbf{x}_i^n = [\textbf{x}_i[1],\cdots,\textbf{x}_i[n]] $. At the TS $t$, the input signal is encoded by
	\begin{equation}
	\textbf{x}_i[t] = f_i(W_i,\textbf{y}_i^{t-1}), \quad i = 1,2. \label{1S}
	\end{equation}
	A decoder $g_j(\cdot)$ at the receiver $j=1,2$ decodes the output signal $\textbf{y}_j^n$ to an estimated message $\widehat{W}_j$, which is given by
	\begin{equation}
	\widehat{W}_j = g_j(\textbf{H}_1^n,\textbf{H}_2^n,\textbf{y}_j^n), \quad j =1,2,
	\end{equation}
	where we assume that the two receivers have perfect CSI.
	
	 According to \cite{30}, a SDoF tuple $(d_1,d_2)$ is 
	 achievable if there exists a secure code satisfying the following reliability constraint:
	\begin{subequations}
			\begin{eqnarray} 
			\lim_{n \rightarrow {\cal{1}}} \Pr[W_j \ne \widehat{W}_j] = 0, \quad j = 1,2,  \label{R1} \\
				\lim_{\text{SNR} \rightarrow {\cal{1}}}    \lim_{n \rightarrow {\cal{1}}} \frac{\log|W_j(n,\text{SNR})|}{n\log \text{SNR}} \ge d_j, \quad j = 1,2, \label{RR1}
		\end{eqnarray}
	\end{subequations}
	and the following security constraint:
	\begin{subequations}
		\begin{eqnarray}
		\lim_{\text{SNR} \rightarrow {\cal{1}}}    \lim_{n \rightarrow {\cal{1}}} \frac{I(W_1;\textbf{y}_2^n)}{n\log \text{SNR}}  = 0, \label{S1} \\
		\lim_{\text{SNR} \rightarrow {\cal{1}}}    \lim_{n \rightarrow {\cal{1}}} \frac{I(W_2;\textbf{y}_1^n)}{n\log \text{SNR}}  = 0. \label{S2}
		\end{eqnarray}
	\end{subequations}

\subsection{Sum-SDoF Problem}

For the $(M,M,N,N)$ MIMO interference channel with confidential messages, we consider a following problem: How to achieve the sum-SDoF, namely the maximal sum of SDoF tuple, when local output feedback, reliability, and security constraints are satisfied? Mathematically, the above problem can be written as follows:
\begin{equation}
  \text{SDoF}_{\text{Sum}}^{\text{L-OF}} = \max_{(d_1,d_2)} \,\,\, d_1+d_2,  
\end{equation}
where $(d_1,d_2)$ are under the constraints \eqref{1S}, \eqref{R1}, \eqref{RR1}, \eqref{S1}, and \eqref{S2}. Specifically, the constraint \eqref{1S} indicates that transmit signals are subject to local output feedback. The constraints \eqref{R1} and \eqref{RR1} indicate that the message can be reliably decoded at the receiver. The constraints \eqref{S1} and \eqref{S2} indicate that there is zero information leakage. In the following, we will present the transmission schemes, and show that the sum-SDoF lower bound achieved by the schemes is the sum-SDoF for $M\le N/2$, $M=N$, and $2N\le M$ antenna configurations.

\section{Transmission Scheme Design}

The interference decoding scheme is presented in Sub-section IV-B. The interference alignment scheme is proposed in Sub-section IV-C. The interference alignment scheme with $M=2N$ is given in Sub-section IV-D.

    For these three schemes, the phase duration  $\tau_1,\tau_2$, and $\tau_3$ are undetermined initially, and then designed to maximize the sum-SDoF lower bound achieved by transmission schemes under decoding and security constraints. The number of transmitted data symbols are given by $2N\tau_2$, $2M\tau_2$, and $4N\tau_2$, for $N/2 < M \le N$, $N < M \le 2N$, and $2N < M$ antenna configurations, respectively. The derived optimal phase duration is summarized in Table \ref{Tab1}.

	\begin{table}[!h]
		\centering
		\caption{Optimal Phase Duration $\tau_1$, $\tau_2$, and $\tau_3$}
		\begin{tabular}{c|c|c|c}
			\hline 
		 	$\backslash$ & $N/2 < M \le N$ & $N < M \le 2N$  & $2N < M$  \\ \hline 
			$\tau_1$ & $N$ & $N$ & $N$   \\  \hline
	 	$\tau_2$ & $2M - N$ & $N$ & $N$   \\  \hline
			$\tau_3$ & $\backslash$ & $M-N$ & $N$   \\  \hline 
		\end{tabular}
		\label{Tab1}
	\end{table}

\subsection{$M \le N/2$ Case: Keep Two Transmitters Silent}

For $M \le N/2$ antenna configurations, all the AN symbols transmitted by two transmitters can be immediately decoded by the receivers. Consequently, we need to keep two transmitters silent. Thus, the sum-SDoF lower bound is $0$.  
	
\subsection{$N/2 < M \le N$ Case: The Interference Decoding Scheme}

For $N/2 < M \le N$ antenna configurations, we propose an interference decoding scheme, since each receiver can decode the transmitted data signals from transmitters 1 and 2 immediately, if we carefully arrange the transmission. In the following, we first provide an illustrative example of the interference decoding scheme and then introduce the general implementation of the scheme.

\subsubsection{An Example}

To help the readers to establish the sense of our transmission scheme,
we begin this part with an illustrative example, where we set $M=2$ and $N=3$. We will show that we can securely deliver a total of $6$ data symbols over $5$ TSs, i.e., $6/5$ sum-SDoF lower bound. The transmission scheme has three phases, and is elaborated as follows:

Phase-I has $3$ TSs and is used to send AN signals. In each TS of Phase-I, each transmitter sends $2$ fresh AN symbols simultaneously. After Phase-I transmission, each receiver has $9$ equations about $12$ AN symbols. Thus, none of AN symbols can be decoded. After Phase-I transmission, the receivers 1 and 2's output signals of Phase-I are fed back to receivers 1 and 2, respectively.

Phase-II has $1$ TS and is used to deliver $2$ fresh data symbols for transmitter 1 and $1$ fresh data symbol for transmitter 2.  In each TS of Phase-II, transmitter 1 sends $2$ fresh data symbols along with the receiver 1's output signals of Phase-I, and transmitter 2 sends $1$ fresh data symbol along with the receiver 2's output signals of Phase-I. Since each receiver has $3$ antennas, it can decode all the received signals. Thus, after Phase-II transmission, receiver 1 can obtain $2$ fresh data symbols and receiver 2 can obtain $1$ fresh data symbols. Meanwhile, receiver 1 acquires extra $1$ equation about $12$ AN symbols, and receiver 2 acquires extra $2$ equations about $12$ AN symbols. 

Phase-III has $1$ TS and is used to deliver $1$ fresh data symbol for transmitter 1 and $2$ fresh data symbols for transmitter 2. In each TS of Phase-III, transmitter 1 sends $1$ fresh data symbol along with the receiver 1's output signals of Phase-I, and transmitter 2 sends $2$ fresh data symbols along with the receiver 2's output signals of Phase-I. Since each receiver has $3$ antennas, it can decode all the received signals. Thus, after Phase-III transmission, receiver 1 can obtain $1$ fresh data symbol and receiver 2 can obtain $2$ fresh data symbols. Meanwhile, receiver 1 acquires extra $2$ equations about $12$ AN symbols, and receiver 2 acquires extra $1$ equation about $12$ AN symbols.

Overall, each receiver obtains $12$ equations about $12$ AN symbols. This implies no extra decoding ability can be used for eavesdropping on the other receiver's data symbols. On the other hand, each receiver can obtain $3$ equations about $3$ desired data symbols from Phase-II and Phase-III transmissions, by subtracting the Phase-I output signals. Therefore, a total of $6$ data symbols are securely delivered over $5$ TSs. 

\subsubsection{The General Scheme}
 
\begin{figure}[t] 
	\centering
	\includegraphics[width=3.5in]{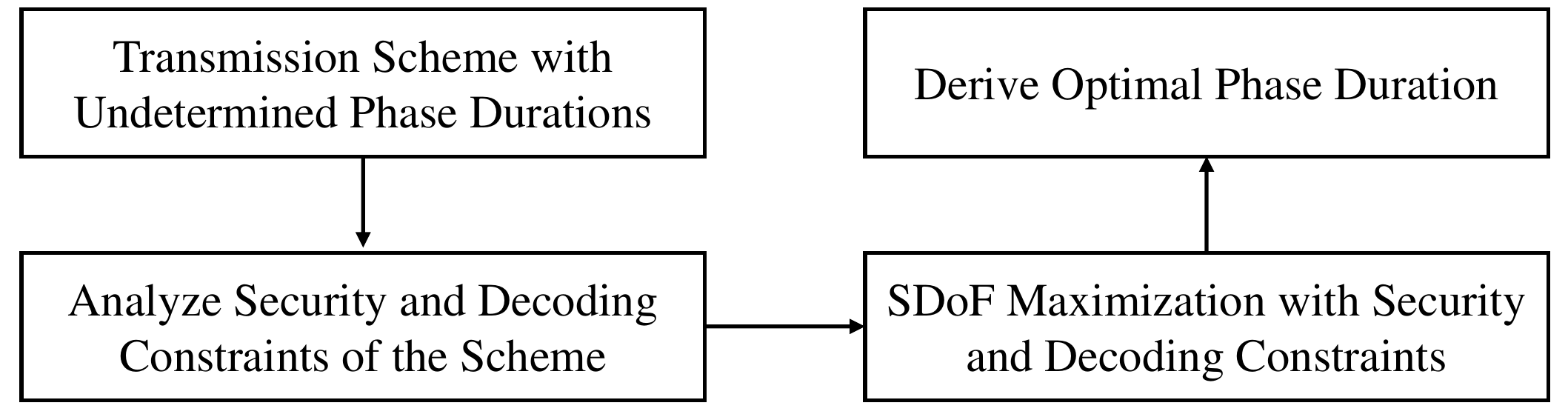}
	\caption{Flow of the proposed solution technique for phase duration design.} \label{Flow}
\end{figure}
\begin{figure}[t]
	\begin{center}
		\includegraphics[width=2.8in]{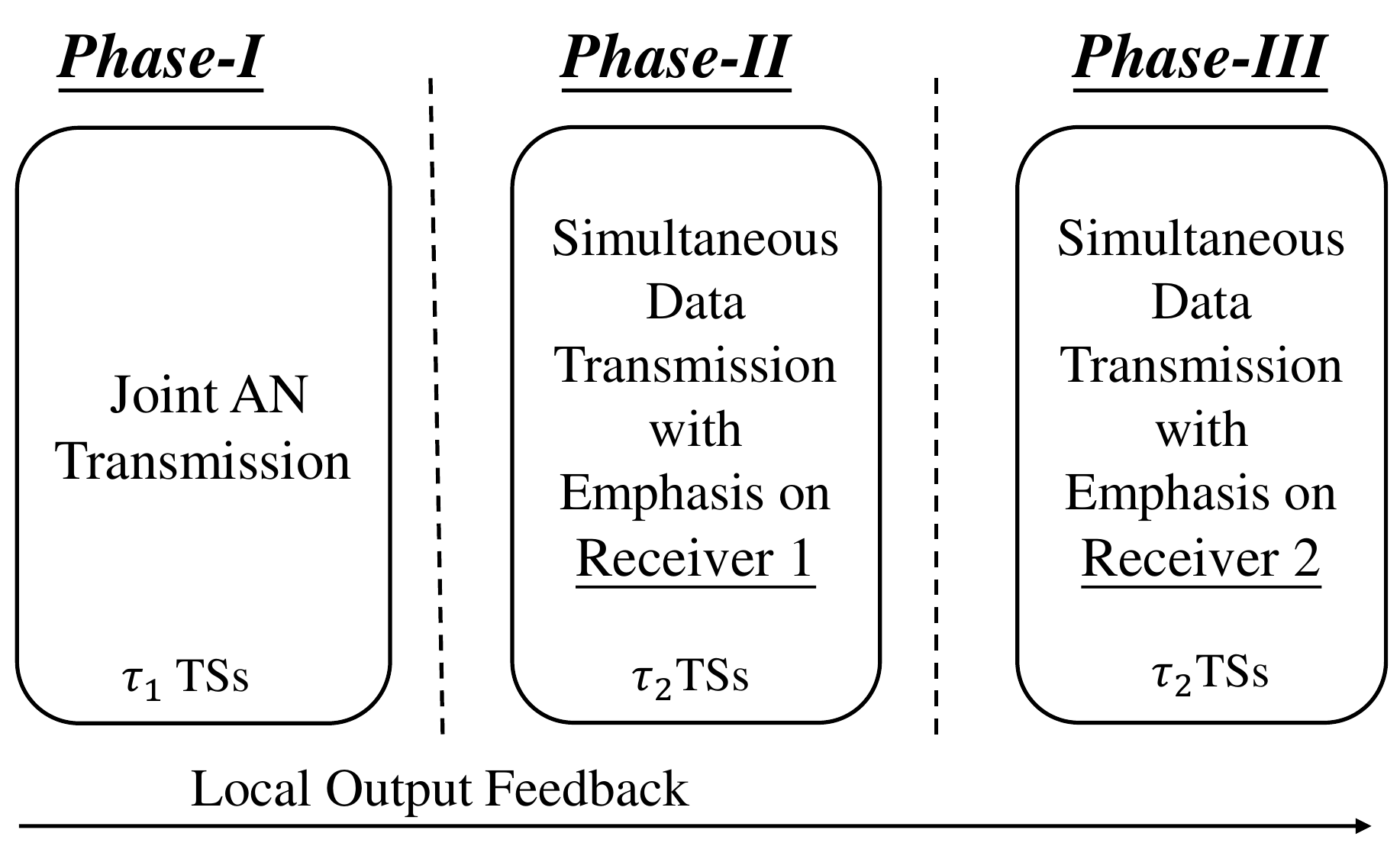}
	\end{center}	
	\caption{Flow-diagram for the interference decoding scheme.} \label{D1}
\end{figure}

To generalize the above example, the major difficulty is how to design the phase duration. This is because, heuristic phase duration assignment cannot guarantee any optimality. To tackle this problem, we first assume the undermined phase duration for the general scheme and then derive the security and decoding constraints with the undermined phase duration. Finally, we determine the optimal phase duration by maximizing the sum-SDoF lower bound with the security and decoding constraints. The flow of the proposed solution technique is given in Fig. \ref{Flow}. In the following, the general transmission scheme with undetermined phase duration is elaborated below, whose flow-diagram is depicted in Fig. \ref{D1}.

First of all, we define the effective CSI matrices of Phase-I, Phase-II, and Phase-III as follows:
\begin{subequations}
	\begin{eqnarray}
	&& \textbf{H}_{i,j}^{\text{I}} = \text{bd} \{\textbf{H}_{i,j}[1], \cdots, \textbf{H}_{i,j}[\tau_1]\}, \nonumber \\
	&& \textbf{H}_{i,j}^{\text{II}} = \text{bd} \{\textbf{H}_{i,j}[\tau_1 + 1], \cdots,  \textbf{H}_{i,j}[\tau_1 + \tau_2]\}, \nonumber \\
	&& \textbf{H}_{i,j}^{\text{III}} = \text{bd} \{\textbf{H}_{i,j}[\tau_1 + \tau_2+1], \cdots,  \textbf{H}_{i,j}[\tau_1 + 2\tau_2]\},  \nonumber
	\end{eqnarray}
\end{subequations}
where $i,j =1,2$.  In addition, we define the following full rank and pre-assigned matrices for compressing the output signals:
${\bf{\Phi}}_1^{a} \in \mathbb{C}^{M\tau_2\times N\tau_1}$,${\bf{\Phi}}_2^{a} \in \mathbb{C}^{(N-M)\tau_2\times N\tau_1}$,${\bf{\Phi}}_1^{b} \in \mathbb{C}^{(N-M)\tau_2\times N\tau_1}$,${\bf{\Phi}}_2^{b} \in \mathbb{C}^{M\tau_2\times N\tau_1}$.
Each matrix can be generated by the standard Gaussian distribution and pre-stored at all transmitters and receivers.

\textit{Phase-I (Joint AN Transmission)}: This phase spans $\tau_1$ TSs with the aim of joint AN symbol transmission. In each TS, transmitter 1 sends $M\tau_1$ fresh AN symbols by $M$ transmit antennas, and transmitter 2 sends $M\tau_1$ fresh AN symbols by $M$ transmit antennas.
As a holistic view, we denote all the AN symbols sent from transmitter 1 by $\textbf{u}_1 \in \mathbb{C}^{M\tau_1}$ and all the AN symbols sent from transmitter 2 by $\textbf{u}_2 \in \mathbb{C}^{M\tau_1}$. Consequently, the transmitted signals of Phase-I are expressed as
\begin{subequations}
	\begin{eqnarray}
	&& \textbf{x}_1^{\text{I}} = \textbf{u}_1,   \\
	&& \textbf{x}_2^{\text{I}} = \textbf{u}_2,
	\end{eqnarray}
\end{subequations}
where we denote the transmitter 1's transmit signals of Phase-I by $\textbf{x}_1^{\text{I}} \in \mathbb{C}^{M\tau_1}$ and the transmitter 2's transmit signals of Phase-I by $\textbf{x}_2^{\text{I}} \in \mathbb{C}^{M\tau_1}$. At the receivers, the output signals are written by
\begin{subequations}
	\begin{eqnarray}
	&& \textbf{y}_1^{\text{I}} = \textbf{H}_{1,1}^{\text{I}} \textbf{u}_1 +  \textbf{H}_{2,1}^{\text{I}} \textbf{u}_2  + \textbf{z}_1^\text{I}, \\
	&& \textbf{y}_2^{\text{I}} = \textbf{H}_{1,2}^{\text{I}} \textbf{u}_1 +  \textbf{H}_{2,2}^{\text{I}} \textbf{u}_2 + \textbf{z}_2^\text{I},
	\end{eqnarray}
\end{subequations}
where we denote the receiver 1's output signals of Phase-I by $\textbf{y}_1^{\text{I}} \in \mathbb{C}^{N\tau_1}$, the receiver 2's output signals of Phase-I by $\textbf{y}_2^{\text{I}} \in \mathbb{C}^{N\tau_1}$, and the AWGN at receivers 1 and 2 by $\textbf{z}_1^\text{I} \in \mathbb{C}^{N\tau_1}$ and $\textbf{z}_2^\text{I} \in \mathbb{C}^{N\tau_1}$, respectively. Besides, the dimension of $\textbf{H}_{i,j}^{\text{I}},i,j=1,2$ is  $N\tau_1 \times M\tau_1$. After Phase-I transmission, the output signals $\textbf{y}_1^\text{I}$ and $\textbf{y}_2^\text{I}$  are fed back to receivers 1 and 2, respectively.

\textit{Phase-II (Simultaneous Data Transmission with Emphasis on Receiver 1)}: This phase spans $\tau_2$ TSs with the aim of fresh data transmission for receivers 1 and 2. In each TS, $M$ fresh data symbols are sent from transmitter 1, and $N-M$ fresh data symbols are sent from transmitter 2. As a holistic view, we denote all the fresh data symbols for receiver 1 by $\textbf{s}_1^{a} \in \mathbb{C}^{M\tau_2}$ and all the fresh data symbols for receiver 2 by $\textbf{s}_2^{a} \in \mathbb{C}^{(N-M)\tau_2}$. To match the dimension of data symbols, we also compress the output signals into ${\bf{\Phi}}_1^{a}\textbf{y}_{1}^{\text{I}} \in \mathbb{C}^{M\tau_2}$ and ${\bf{\Phi}}_2^{a}\textbf{y}_{2}^{\text{I}} \in \mathbb{C}^{(N-M)\tau_2}$. For security, the transmitted signals of Phase-II are designed by
\begin{subequations}
	\begin{eqnarray}
	&& \textbf{x}_1^{\text{II}} = \textbf{s}_1^{a} + {\bf{\Phi}}_1^{a}\textbf{y}_{1}^{\text{I}},    \\
	&& \textbf{x}_2^{\text{II}} = \textbf{s}_2^{a} + {\bf{\Phi}}_2^{a}\textbf{y}_{2}^{\text{I}},
	\end{eqnarray}
\end{subequations}
where we denote the transmitter 1's transmit signals of Phase-II by $\textbf{x}_1^{\text{II}} \in \mathbb{C}^{M\tau_2}$ and the transmitter 2's transmit signals of Phase-II by $\textbf{x}_2^{\text{II}} \in \mathbb{C}^{(N-M)\tau_2}$. At the receivers, the output signals are written by
\begin{subequations}
	\begin{eqnarray}
	&& \textbf{y}_1^{\text{II}} = \begin{bmatrix}
	\textbf{H}_{1,1}^{\text{II}} &  \textbf{H}_{2,1}^{\text{II}}
	\end{bmatrix}  \begin{bmatrix}
	\textbf{s}_1^{a} + {\bf{\Phi}}_1^{a}\textbf{y}_{1}^{\text{I}} \\    
	\textbf{s}_2^{a} + {\bf{\Phi}}_2^{a}\textbf{y}_{2}^{\text{I}}
	\end{bmatrix} + \textbf{z}_1^\text{II},\\
	&& \textbf{y}_2^{\text{II}} = \begin{bmatrix}
	\textbf{H}_{1,2}^{\text{II}} &  \textbf{H}_{2,2}^{\text{II}}
	\end{bmatrix} \begin{bmatrix}
	\textbf{s}_1^{a} + {\bf{\Phi}}_1^{a}\textbf{y}_{1}^{\text{I}} \\    
	\textbf{s}_2^{a} + {\bf{\Phi}}_2^{a}\textbf{y}_{2}^{\text{I}}
	\end{bmatrix} + \textbf{z}_2^\text{II},
	\end{eqnarray}
\end{subequations}
where we denote the receiver 1's output signals of Phase-II by $\textbf{y}_1^{\text{II}} \in \mathbb{C}^{N\tau_2}$, the receiver 2's output signals of Phase-II by $\textbf{y}_2^{\text{II}} \in \mathbb{C}^{N\tau_2}$, and the AWGN at receivers 1 and 2 by $\textbf{z}_1^\text{II}$ and $\textbf{z}_2^\text{II}$, respectively. Besides, the dimension of $\textbf{H}_{1,j}^{\text{II}},j=1,2$ is  $N\tau_2 \times M\tau_2$ and the dimension of $\textbf{H}_{2,j}^{\text{II}},j=1,2$  is $N\tau_2 \times (N-M)\tau_2$.

\textit{Phase-III (Simultaneous Data Transmission with Emphasis on Receiver 2)}:  This phase spans $\tau_2$ TSs with the aim of fresh data transmission for receivers 1 and 2. In each TS, $N-M$ fresh data symbols are sent from transmitter 1, and $M$ fresh data symbols are sent from transmitter 2. As a holistic view, we denote all the fresh data symbols for receiver 1 by $\textbf{s}_1^{b} \in \mathbb{C}^{(N-M)\tau_2}$ and all the fresh data symbols for receiver 2 by $\textbf{s}_2^{b} \in \mathbb{C}^{M\tau_2}$. To match the dimension of data symbols, we also compress the output signals into ${\bf{\Phi}}_1^{b}\textbf{y}_{1}^{\text{I}} \in \mathbb{C}^{(N-M)\tau_2}$ and ${\bf{\Phi}}_2^{b}\textbf{y}_{2}^{\text{I}} \in \mathbb{C}^{M\tau_2}$. For security, the transmitted signals of Phase-III are designed by
\begin{subequations}
	\begin{eqnarray}
	&& \textbf{x}_1^{\text{III}} = \textbf{s}_1^{b} + {\bf{\Phi}}_1^{b}\textbf{y}_{1}^{\text{I}},    \\
	&& \textbf{x}_2^{\text{III}} = \textbf{s}_2^{b} + {\bf{\Phi}}_2^{b}\textbf{y}_{2}^{\text{I}},
	\end{eqnarray}
\end{subequations}
where we denote the transmitter 1's transmit signals of Phase-III by $\textbf{x}_1^{\text{III}} \in \mathbb{C}^{(N-M)\tau_2}$ and the transmitter 2's transmit signals of Phase-III by $\textbf{x}_2^{\text{III}} \in \mathbb{C}^{M\tau_2}$. At the receivers, the output signals are written by
\begin{subequations}
	\begin{eqnarray}
	&&\textbf{y}_1^{\text{III}} = \begin{bmatrix}
	\textbf{H}_{1,1}^{\text{III}} &  \textbf{H}_{2,1}^{\text{III}}
	\end{bmatrix}  \begin{bmatrix}
	\textbf{s}_1^{b} + {\bf{\Phi}}_1^{b}\textbf{y}_{1}^{\text{I}} \\    
	\textbf{s}_2^{b} + {\bf{\Phi}}_2^{b}\textbf{y}_{2}^{\text{I}}
	\end{bmatrix} + \textbf{z}_1^\text{III}, \\
	&&\textbf{y}_2^{\text{III}} = \begin{bmatrix}
	\textbf{H}_{1,2}^{\text{III}} &  \textbf{H}_{2,2}^{\text{III}}
	\end{bmatrix} \begin{bmatrix}
	\textbf{s}_1^{b} + {\bf{\Phi}}_1^{b}\textbf{y}_{1}^{\text{I}} \\    
	\textbf{s}_2^{b} + {\bf{\Phi}}_2^{b}\textbf{y}_{2}^{\text{I}}
	\end{bmatrix} + \textbf{z}_2^\text{III},
	\end{eqnarray}
\end{subequations}
where we denote the receiver 1's output signals of Phase-III by $\textbf{y}_1^{\text{III}} \in \mathbb{C}^{N\tau_2}$, and  the receiver 2's output signals of Phase-III by $\textbf{y}_2^{\text{III}} \in \mathbb{C}^{N\tau_2}$, and the AWGN at receivers 1 and 2 by $\textbf{z}_1^\text{III} \in \mathbb{C}^{N\tau_2}$ and $\textbf{z}_2^\text{III} \in \mathbb{C}^{N\tau_2}$, respectively.
Besides, the dimension of $\textbf{H}_{1,j}^{\text{III}},j=1,2$  is $N\tau_2 \times (N-M)\tau_2$ and the dimension of $\textbf{H}_{2,j}^{\text{III}},j=1,2$ is $N\tau_2 \times M\tau_2$.

The major concern of the transmission scheme is that we must ensure the security and decodability after transmission. To this end, we perform the security and decoding analysis in the following. Specifically, the key of security analysis is to verify the information leakage to the other receiver is zero, i.e., $I(\textbf{s}_1^{a}, \textbf{s}_1^{b};\textbf{y}_2|\textbf{s}_2^{a}, \textbf{s}_2^{b})=0$. However, the direct verification of zero information leakage is difficult. Hence, we resort to data processing inequality to transform it into a tractable one.  

\textit{Security Analysis}: To ensure security, zero information leakage to the eavesdropper should be guaranteed. For simplicity, we define $\textbf{y}_2 = [\textbf{y}_2^{\text{I}}; \textbf{y}_2^{\text{II}};\textbf{y}_2^{\text{III}}]$ and $\textbf{u} = [\textbf{u}_1;\textbf{u}_2]$.
The verification of information leakage $I(\textbf{s}_1^{a}, \textbf{s}_1^{b};\textbf{y}_2|\textbf{s}_2^{a}, \textbf{s}_2^{b})  = 0$ is presented in  \eqref{Q1}, shown at the top of next page, 
where the reason of each critical step is given as follows:
\begin{enumerate}[(a)]
	\item Data processing inequality for Markov chain $(\textbf{s}_1^{a},  \textbf{s}_1^{b}, \textbf{u}) \rightarrow ([\textbf{H}^{\text{I}}_{1,2}, \textbf{H}^{\text{I}}_{2,2}]\textbf{u}, \textbf{H}_{1,2}^\text{II}\textbf{s}_1^{a} + \textbf{H}_{1,2}^{\text{II}}{\bf{\Phi}}_1^{a}[\textbf{H}_{1,1}^{\text{I}}, \textbf{H}_{2,1}^{\text{I}}]\textbf{u}$, $\textbf{H}_{1,2}^\text{III}\textbf{s}_1^{b} + \textbf{H}_{1,2}^{\text{III}}{\bf{\Phi}}_1^{b}[\textbf{H}_{1,1}^{\text{I}}, \textbf{H}_{2,1}^{\text{I}}]\textbf{u}) \rightarrow \textbf{y}_2$.
	\item When input is circularly symmetric complex Gaussian, according to \cite{100}, rewriting into $\log\text{det}(\textbf{I} + \text{SNR}\textbf{A}\textbf{A}^H) - \log\text{det}(\textbf{I} + \text{SNR}\textbf{B}\textbf{B}^H)$, and using  Lemma 2 in \cite{20}.
	\item The rank of matrix $\textbf{A}$ is $N(\tau_1 + \tau_2)$. The rank of matrix $\textbf{B}$  is $\min\{N(\tau_1+\tau_2),2M\tau_1\}$, whose reason is given in Appendix A.
\end{enumerate}
Due to the symmetry of antenna configuration at receivers, the security analysis of the other receiver is similar. Therefore, to ensure security, according to \eqref{Q1}, it should be followed that 
\begin{equation}
\label{Z3}
N(\tau_1 + \tau_2) \le 2M\tau_1,
\end{equation}
which is the security constraint. Next, we perform the decoding analysis.

\begin{figure*}[!ht]    
	\begin{eqnarray}
		&& I(\textbf{s}_1^{a}, \textbf{s}_1^{b};\textbf{y}_2|\textbf{s}_2^{a}, \textbf{s}_2^{b})   = I( \textbf{s}_1^{a},  \textbf{s}_1^{b}, \textbf{u};\textbf{y}_2|\textbf{s}_2^{a}, \textbf{s}_2^{b}) - I(\textbf{u};\textbf{y}_2| \textbf{s}_1^{a}, \textbf{s}_1^{b},\textbf{s}_2^{a}, \textbf{s}_2^{b}) \nonumber \\
		&& \overset{(a)}{\le} I([\textbf{H}^{\text{I}}_{1,2}, \textbf{H}^{\text{I}}_{2,2}]\textbf{u}, \textbf{H}_{1,2}^\text{II}\textbf{s}_1^{a} + \textbf{H}_{1,2}^{\text{II}}{\bf{\Phi}}_1^{a}[\textbf{H}_{1,1}^{\text{I}}, \textbf{H}_{2,1}^{\text{I}}]\textbf{u}, \textbf{H}_{1,2}^\text{III}\textbf{s}_1^{b} + \textbf{H}_{1,2}^{\text{III}}{\bf{\Phi}}_1^{b}[\textbf{H}_{1,1}^{\text{I}}, \textbf{H}_{2,1}^{\text{I}}]\textbf{u};\textbf{y}_2|\textbf{s}_2^{a}, \textbf{s}_2^{b}) \nonumber \\
		&& - I(\textbf{u};\textbf{y}_2| \textbf{s}_1^{a}, \textbf{s}_1^{b},\textbf{s}_2^{a}, \textbf{s}_2^{b}) \nonumber \\
		&& \underset{\text{SNR}\rightarrow {\cal{1}}}{\overset{(b)}{=}}  \text{rk}  \left\{
		\underbrace{\begin{bmatrix}
				\textbf{I}_{N\tau_1} &  \textbf{0} & \textbf{0}\\
				\textbf{0} &  \textbf{I}_{M\tau_2}  & \textbf{0} \\
				\textbf{0} & \textbf{0} &  \textbf{I}_{(N-M)\tau_2}
		\end{bmatrix}}_{\textbf{A}}\right\} \log \text{SNR} \nonumber \\
		&&  - \text{rk} \left\{\underbrace{\begin{bmatrix}
				\textbf{H}_{1,2}^{\text{I}}  & \textbf{H}_{2,2}^{\text{I}}  \nonumber \\
				\textbf{H}_{1,2}^{\text{II}}{\bf{\Phi}}_1^{a}\textbf{H}_{1,1}^{\text{I}}  + \textbf{H}_{2,2}^{\text{II}}{\bf{\Phi}}_2^{a}\textbf{H}_{1,2}^{\text{I}} &  \textbf{H}_{1,2}^{\text{II}}{\bf{\Phi}}_1^{a}\textbf{H}_{2,1}^{\text{I}} + \textbf{H}_{2,2}^{\text{II}}{\bf{\Phi}}_2^{a}\textbf{H}_{2,2}^{\text{I}}\nonumber \\
				\textbf{H}_{1,2}^{\text{III}}{\bf{\Phi}}_1^{b}\textbf{H}_{1,1}^{\text{I}} + \textbf{H}_{2,2}^{\text{III}}{\bf{\Phi}}_2^{b}\textbf{H}_{1,2}^{\text{I}}  &  \textbf{H}_{1,2}^{\text{III}}{\bf{\Phi}}_1^{b}\textbf{H}_{2,1}^{\text{I}} +
				\textbf{H}_{2,2}^{\text{III}}{\bf{\Phi}}_2^{b}\textbf{H}_{2,2}^{\text{I}}
		\end{bmatrix}}_{\textbf{B}}\right\} \log \text{SNR} \nonumber \\
		&& \overset{(c)}{=} N(\tau_1 + \tau_2)\log \text{SNR} - \min\{N(\tau_1+\tau_2),2M\tau_1\} \log \text{SNR}  = 0. \label{Q1}
	\end{eqnarray}
	\hrule
\end{figure*}

\begin{figure*}[ht]
	\begin{subequations}
			\begin{eqnarray}
			&& \begin{bmatrix}
				\textbf{y}_1^\text{II} - \textbf{H}_{1,1}^\text{II}{\bf{\Phi}}_1^a\textbf{y}_1^\text{I}\\
				\textbf{y}_1^\text{III} - \textbf{H}_{1,1}^\text{III}{\bf{\Phi}}_1^b\textbf{y}_1^\text{I}
			\end{bmatrix}   = \underbrace{\begin{bmatrix}
					\textbf{H}_{1,1}^\text{II} & \textbf{H}_{2,1}^\text{II} & \textbf{0} & \textbf{0} \\
					\textbf{0} & \textbf{0} & \textbf{H}_{1,1}^\text{III} & \textbf{H}_{2,1}^\text{III}
			\end{bmatrix}}_{\textbf{H}_1}
			\begin{bmatrix}
				\textbf{s}_1^{a}  \\    
				\textbf{s}_2^{a} + {\bf{\Phi}}_2^{a}\textbf{y}_{2}^{\text{I}} \\
				\textbf{s}_1^{b}   \\    
				\textbf{s}_2^{b} + {\bf{\Phi}}_2^{b}\textbf{y}_{2}^{\text{I}}     \end{bmatrix}  + \overline{\textbf{z}}_1, \label{V1} \\
			&& \begin{bmatrix}
				\textbf{y}_2^\text{II} - \textbf{H}_{2,2}^\text{II}{\bf{\Phi}}_2^a\textbf{y}_2^\text{I}\\
				\textbf{y}_2^\text{III} - \textbf{H}_{2,2}^\text{III}{\bf{\Phi}}_2^b\textbf{y}_2^\text{I}
			\end{bmatrix}
			= \underbrace{\begin{bmatrix}
					\textbf{H}_{1,2}^\text{II} & \textbf{H}_{2,2}^\text{II} & \textbf{0} & \textbf{0} \\
					\textbf{0} & \textbf{0}  & \textbf{H}_{1,2}^\text{III} & \textbf{H}_{2,2}^\text{III}
			\end{bmatrix}}_{\textbf{H}_2}\begin{bmatrix}
				\textbf{s}_1^{a}  + {\bf{\Phi}}_1^{a}\textbf{y}_{1}^{\text{I}}  \\    
				\textbf{s}_2^{a}  \\
				\textbf{s}_1^{b}   + {\bf{\Phi}}_1^{b}\textbf{y}_{1}^{\text{I}}  \\    
				\textbf{s}_2^{b}
			\end{bmatrix} + \overline{\textbf{z}}_2. \label{V2}
		\end{eqnarray} 
	\end{subequations}	  
	\hrule  
\end{figure*}

\textit{Decoding Analysis}: The decoding equation for receiver 1 and 2 is given in \eqref{V1} and \eqref{V2}, respectively, where $\overline{\textbf{z}}_1$ and $\overline{\textbf{z}}_2$ are the AWGN vectors, and $\textbf{H}_1$ and $\textbf{H}_2$ are effective decoding channel. Accordingly, the information rate for receiver 1 is evaluated as follows:
\begin{eqnarray}
&& I(\textbf{s}_1^{a}, \textbf{s}_1^{b};\textbf{y}_1|\textbf{y}_{1}^{\text{I}})   = I(\textbf{s}_1^{a}, \textbf{s}_1^{b},  \textbf{s}_2^{a} +  {\bf{\Phi}}_2^{a}\textbf{y}_{2}^{\text{I}},    \textbf{s}_2^{b} +  {\bf{\Phi}}_2^{b}\textbf{y}_{2}^{\text{I}};\textbf{y}_1|\textbf{y}_{1}^{\text{I}}) \nonumber \\
&& - I( \textbf{s}_2^{a} +  {\bf{\Phi}}_2^{a}\textbf{y}_{2}^{\text{I}},    \textbf{s}_2^{b} +  {\bf{\Phi}}_2^{b}\textbf{y}_{2}^{\text{I}};\textbf{y}_1|\textbf{s}_1^{a}, \textbf{s}_1^{b}, \textbf{y}_{1}^{\text{I}}) \nonumber \\
&& \underset{\text{SNR}\rightarrow {\cal{1}}}{\overset{(a)}{=}} \text{rk}\{\textbf{H}_1\} \log \text{SNR}   - \text{rk} \{\text{bd}\{\textbf{H}_{2,1}^{\text{II}},\textbf{H}_{2,1}^{\text{III}}\}\} \log \text{SNR} \nonumber \\
&& = 2N \log \text{SNR} - N \log \text{SNR}  = N \log \text{SNR}, \label{J1}
\end{eqnarray}
where the reason of critical step is given as follows:
\begin{enumerate}[(a)]
	\item  When input is circularly symmetric complex Gaussian, according to \cite{100}, rewriting into $\log \text{det}(\textbf{I}+\text{SNR}\textbf{H}_1\textbf{H}_1^H) - \log \text{det}(\textbf{I}+\text{SNR}\text{bd}\{\textbf{H}_{2,1}^{\text{II}},\textbf{H}_{2,1}^{\text{III}}\}\text{bd}\{\textbf{H}_{2,1}^{\text{II}},\textbf{H}_{2,1}^{\text{III}}\}^H)$, and using  Lemma 2 in \cite{20}.
\end{enumerate}
Therefore, $N$ data symbols desired by receiver 1 can be decoded. Due to the symmetry of antenna configuration at receivers, we have $I(\textbf{s}_2^{a}, \textbf{s}_2^{b};\textbf{y}_2) = N \log \text{SNR}$ as well. Consequently, the decodability of $N$ data symbols desired by receiver 2 can be guaranteed. With the security and decoding constraints, we are able to optimize the phase duration.

\textit{Optimal Phase Duration}: We propose to optimize the phase duration by maximizing the sum-SDoF lower bound achieved by our scheme with the security constraint, as the decodability can be always guaranteed,. In the general scheme, we securely deliver $2N\tau_1$ data symbols over $\tau_1 + 2\tau_2$ TSs. Hence, we formulate a linear-fractional optimization problem, given by
\begin{eqnarray}
\max_{\tau_1,\tau_2} & \dfrac{2N\tau_2}{\tau_1 + 2\tau_2} \\
\text{s.t.} & N(\tau_1 + \tau_2) \le 2M\tau_1.\nonumber
\end{eqnarray}
To solve the problem, we propose to re-formulate it as follows:
\begin{eqnarray}
\max_{\tau_1/\tau_2} & \dfrac{2N}{\tau_1/\tau_2 + 2} \\
\text{s.t.} & N/(2M-N) \le \tau_1/\tau_2.\nonumber
\end{eqnarray}
Since the objective function is inversely proportional to $\tau_1/\tau_2$, the optimal solution is obtained when $\tau_1/\tau_2 = N/(2M-N)$. Therefore, for the original problem, the optimal value is $2N(2M-N)/(4M-N)$ by setting $\tau_1 = N$ and $\tau_2 = 2M - N$.
This shows that the following sum-SDoF lower bound is attainable: 
\begin{equation} \label{ID}
  \text{SDoF}_{\text{Sum}}^{\text{L-OF}} \ge \frac{2N(2M-N)}{4M-N}, \quad \text{if}\,\, N/2 < M \le N.
\end{equation}

\subsection{$N < M \le 2N$ Case: The Interference Alignment Scheme}

For $N < M \le 2N$ antenna configurations, we design an interference alignment scheme, since the number of interfering equations can compensate for the lacking equations for decoding the data symbols, if we carefully arrange the transmission. In the following, we first provide an illustrative example of the interference alignment scheme and then introduce the general implementation of the scheme.

\subsubsection{An Example} 

To help the readers to establish the sense of our transmission scheme,
we begin this part with an illustrative example, where we set $M=3$ and $N=2$. We will show that we can securely deliver a total of $12$ data symbols over $7$ TSs, i.e., $12/7$ sum-SDoF lower bound. The transmission scheme has four phases, and is elaborated as follows:

Phase-I has $2$ TSs and is used to send AN signals with $2$ transmit antennas. In each TS of Phase-I, each transmitter sends $3$ fresh AN symbols simultaneously. After Phase-I transmission, each receiver has $4$ equations about $8$ AN symbols. Thus, none of AN symbols can be decoded. After Phase-I transmission, the receivers 1 and 2's output signals of Phase-I are fed back to receivers 1 and 2, respectively.

Phase-II has $2$ TSs and is used to deliver $6$ fresh data symbols for transmitter 1.  In each TS of Phase-II, transmitter 1 sends $3$ fresh data symbols along with the receiver 1's output signals of Phase-I, and transmitter 2 does not send any signals. Since receiver 1 has $2$ antennas, it cannot decode the fresh data symbols. Meanwhile, receiver 1 does not acquire new equations about $8$ AN symbols, and receiver 2 acquires extra $4$ equations about $8$ AN symbols. After Phase-II transmission, the receiver 2's output signals of Phase-II are fed back to the receiver 2.

Phase-III has $2$ TSs and is used to deliver $6$ fresh data symbols for transmitter 2.  In each TS of Phase-III, transmitter 2 sends $3$ fresh data symbols along with the receiver 2's output signals of Phase-I, and transmitter 1 does not send any signals. Since receiver 2 has $2$ antennas, it cannot decode the fresh data symbols. Meanwhile, receiver 2 does not acquire new equations about $8$ AN symbols, and receiver 1 acquires extra $4$ equations about $8$ AN symbols. After Phase-III transmission, the receiver 1's output signals of Phase-III are fed back to the receiver 1.

Phase-IV has $1$ TS and is used to simultaneously provide $1$ equation about data symbols for transmitters 1 and 2.  In Phase-IV, transmitter 1 sends the receiver 1's output signals of Phase-III, and transmitter 2 sends the receiver 2's output signals of Phase-II. At the receiver 1, it can obtain $1$ equation about the data symbols, by subtracting the receiver 1's output signals of Phase-III from the received signal.  At the receiver 2, it can obtain $1$ equation about the data symbols, by subtracting the receiver 2's output signals of Phase-II from the received signal. Meanwhile, receivers 1 and 2 do not acquire new equations about $8$ AN symbols.

Overall, each receiver obtains $8$ equations about $8$ AN symbols. This implies no extra decoding ability can be used for eavesdropping on the other receiver's data symbols. On the other hand, each receiver can obtain $6$ equations about $6$ desired data symbols. Therefore, a total of $12$ data symbols are securely delivered over $7$ TSs.

\begin{figure}[t]
	\begin{center}
		\includegraphics[width=3.2in]{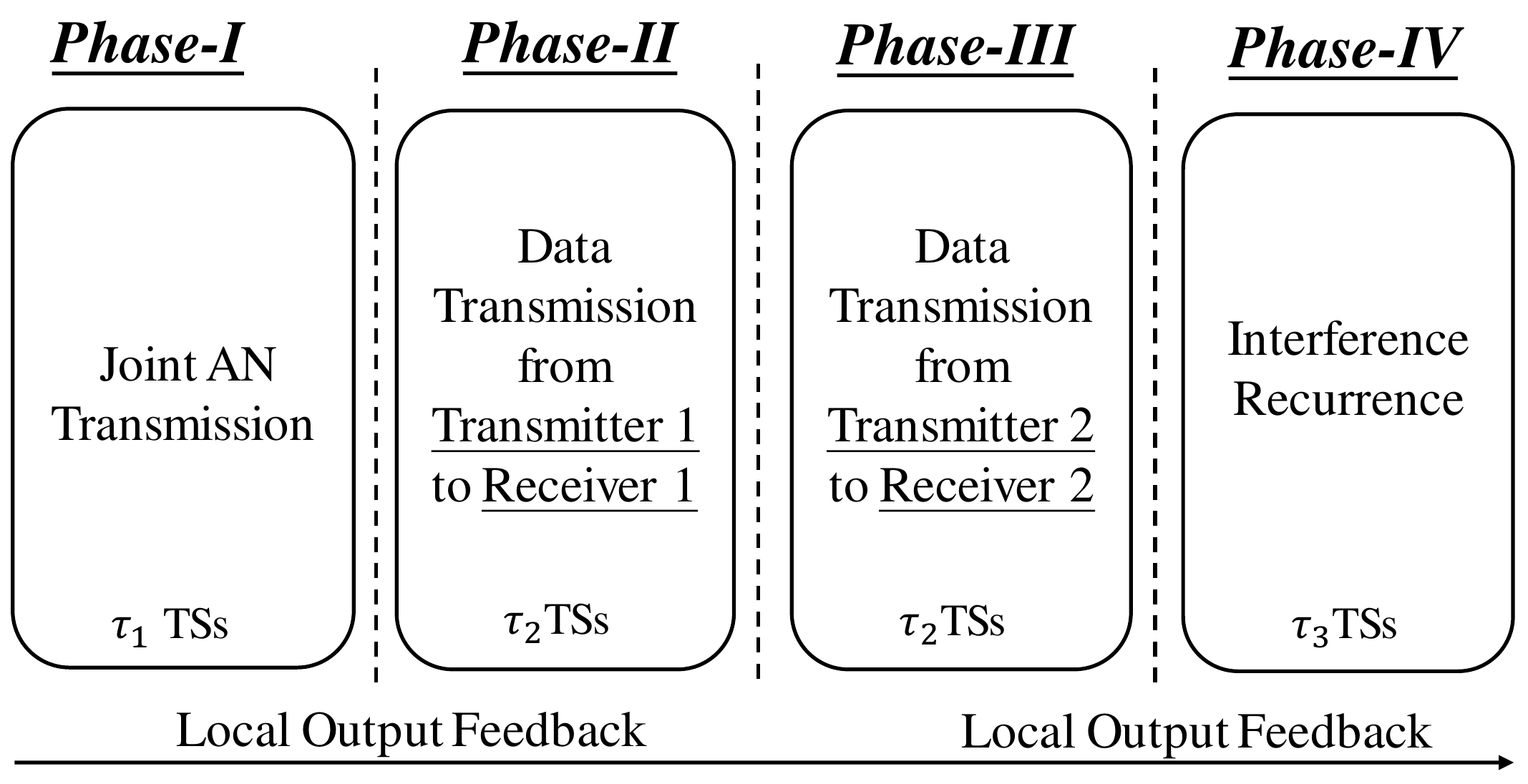}
	\end{center}	
	\caption{Flow-diagram of the interference alignment scheme.} \label{D2}
\end{figure}

\subsubsection{The General Scheme}
To generalize the above example, we need to solve the problem of assigning the phase duration. To tackle this problem, we adopt the design technique presented in sub-section-A. 
In the following, the general transmission scheme with undetermined phase duration is elaborated below, whose flow-diagram is given in Fig. \ref{D2}.

First of all, we define the effective CSI matrices of Phase-I, Phase-II, Phase-III, and Phase-IV as follows:
\begin{subequations}
	\begin{eqnarray}
	&& \textbf{H}_{i,j}^{\text{I}} = \text{bd} \{\textbf{H}_{i,j}[1], \cdots, \textbf{H}_{i,j}[\tau_1]\}, \nonumber \\
	&& \textbf{H}_{i,j}^{\text{II}} = \text{bd} \{\textbf{H}_{i,j}[\tau_1 + 1], \cdots,  \textbf{H}_{i,j}[\tau_1 + \tau_2]\}, \nonumber \\
	&& \textbf{H}_{i,j}^{\text{III}} = \text{bd} \{\textbf{H}_{i,j}[\tau_1 + \tau_2+1], \cdots,  \textbf{H}_{i,j}[\tau_1 + 2\tau_2]\}, \nonumber \\
	&& \textbf{H}_{i,j}^{\text{IV}} = \text{bd} \{\textbf{H}_{i,j}[\tau_1 + 2\tau_2 + 1], \cdots,  \textbf{H}_{i,j}[\tau_1 + 2\tau_2 + \tau_3]\},   \nonumber
	\end{eqnarray}
\end{subequations}
where $i,j = 1,2$. In addition, we define the following full rank and pre-assigned matrices for shaping the received output at transmitters:
${\bf{\Phi}}  \in \mathbb{C}^{M\tau_2 \times N\tau_1} 
$, ${\bf{\Theta}}  \in \mathbb{C}^{N\tau_3 \times N\tau_2}$.
Each matrix can be generated by the standard Gaussian distribution and pre-stored at all transmitters and receivers.

\textit{Phase-I (Joint AN Transmission)}: This phase spans $\tau_1$ TSs with the aim of joint AN symbol transmission. In each TS, transmitter 1 sends $N\tau_1$ fresh AN symbols by $N$ antennas, and transmitter 2 sends $N\tau_1$ fresh AN symbols by $N$ antennas.
As a holistic view, we  denote all the AN symbols sent from transmitter 1 by $\textbf{u}_1 \in \mathbb{C}^{N\tau_1}$ and all the AN symbols sent from transmitter 2 by $\textbf{u}_2 \in \mathbb{C}^{N\tau_1}$. Consequently, the transmitted signals of Phase-I are expressed as
\begin{subequations}
	\begin{eqnarray}
	&& \textbf{x}_1^{\text{I}} = \textbf{u}_1,   \\
	&& \textbf{x}_2^{\text{I}} = \textbf{u}_2,
	\end{eqnarray}
\end{subequations}
where we denote the transmitter 1's transmit signals of Phase-I by $\textbf{x}_1^\text{I} \in \mathbb{C}^{N\tau_1}$ and the transmitter 2's transmit signals of Phase-I by $\textbf{x}_2^\text{I} \in \mathbb{C}^{N\tau_1}$. At the receivers, the output signals are written by
\begin{subequations}
	\begin{eqnarray}
	&& \textbf{y}_1^{\text{I}} = \textbf{H}_{1,1}^{\text{I}} \textbf{u}_1 +  \textbf{H}_{2,1}^{\text{I}} \textbf{u}_2  + \textbf{z}_1^\text{I},\\
	&& \textbf{y}_2^{\text{I}} = \textbf{H}_{1,2}^{\text{I}} \textbf{u}_1 +  \textbf{H}_{2,2}^{\text{I}} \textbf{u}_2+ \textbf{z}_2^\text{I},
	\end{eqnarray}
\end{subequations}
where we denote the receiver 1's output signals of Phase-I by $\textbf{y}_1^{\text{I}} \in \mathbb{C}^{N\tau_1}$, the receiver 2's output signals of Phase-I by $\textbf{y}_2^{\text{I}} \in \mathbb{C}^{N\tau_1}$, and the AWGN at receivers 1 and 2 by $\textbf{z}_1^\text{I} \in \mathbb{C}^{N\tau_1}$ and $\textbf{z}_2^\text{I} \in \mathbb{C}^{N\tau_1}$, respectively. Besides, the dimension of  $\textbf{H}_{i,j}^{\text{I}},i,j=1,2$ is $N\tau_1 \times M\tau_1$. After Phase-I transmission, the output signals $\textbf{y}_1^\text{I}$ and $\textbf{y}_2^\text{I}$  are fed back to receivers 1 and 2, respectively.

\textit{Phase-II (Data Transmission from Transmitter 1 to Receiver 1)}: This phase spans $\tau_2$ TSs with the aim of fresh data symbol transmission for receiver 1. In each TS, $M$ fresh data symbols are sent from transmitter 1, and transmitter 2 does not send any signals. As a holistic view, we denote all the fresh data symbols for receiver 1 by $\textbf{s}_1 \in \mathbb{C}^{M\tau_2}$ and the empty signals by $\textbf{0} \in \mathbb{C}^{M\tau_2}$. To match the dimension of data symbols, we also compress the output signals into ${\bf{\Phi}}\textbf{y}_{1}^{\text{I}} \in \mathbb{C}^{M\tau_2}$. For security, the transmitted signals of Phase-II are designed by
\begin{subequations}
	\begin{eqnarray}
	&&\textbf{x}_1^{\text{II}} = \textbf{s}_1 + {\bf{\Phi}}\textbf{y}_{1}^{\text{I}},    \\
	&& \textbf{x}_2^{\text{II}} = \textbf{0},
	\end{eqnarray}
\end{subequations}
where we denote the transmitter 1's transmit signals of Phase-II by $\textbf{x}_1^\text{II} \in \mathbb{C}^{M\tau_2}$ and the transmitter 2's transmit signals of Phase-II by $\textbf{x}_2^\text{II} \in \mathbb{C}^{M\tau_2}$. At the receivers, the output signals are written by
\begin{subequations}
	\begin{eqnarray}
	&&     \textbf{y}_1^{\text{II}} = \textbf{H}_{1,1}^{\text{II}} \textbf{s}_1 +  \textbf{H}_{1,1}^{\text{II}} {\bf{\Phi}} \textbf{y}_1^{\text{I}} + \textbf{z}_1^\text{II}, \\
	&&     \textbf{y}_2^{\text{II}} = \textbf{H}_{1,2}^{\text{II}} \textbf{s}_1 +  \textbf{H}_{1,2}^{\text{II}} {\bf{\Phi}} \textbf{y}_1^{\text{I}} + \textbf{z}_2^\text{II},
	\end{eqnarray}
\end{subequations}
where we denote the receiver 1's output signals of Phase-II by $\textbf{y}_1^\text{II} \in \mathbb{C}^{N\tau_2}$, the the receiver 2's output signals of Phase-II by $\textbf{y}_2^\text{II} \in \mathbb{C}^{N\tau_2}$, and the AWGN at receivers 1 and 2 by $\textbf{z}_1^\text{II} \in \mathbb{C}^{N\tau_2}$ and $\textbf{z}_2^\text{II} \in \mathbb{C}^{N\tau_2}$, respectively. Besides, the dimension of $\textbf{H}_{i,j}^{\text{II}},i,j=1,2$ is $N\tau_2 \times M\tau_2$.  After Phase-II transmission, the receiver 2's output signals of Phase-II are fed back to the receiver 2.

\textit{Phase-III (Data Transmission from Transmitter 2 to Receiver 2)}: This phase spans $\tau_2$ TSs with the aim of fresh data symbol transmission for receiver 2. In each TS, $M$ fresh data symbols are sent from transmitter 2, and transmitter 1 does not send any signals. As a holistic view, we denote all the fresh data symbols for receiver 2 by $\textbf{s}_2 \in \mathbb{C}^{M\tau_2}$ and the empty signals by $\textbf{0} \in \mathbb{C}^{M\tau_2}$. To match the dimension of data symbols, we also compress the output signals into ${\bf{\Phi}}\textbf{y}_{2}^{\text{I}} \in \mathbb{C}^{M\tau_2}$. For security, the transmitted signals of Phase-III are designed by
\begin{subequations}
	\begin{eqnarray}
	&& \textbf{x}_1^{\text{III}} = \textbf{0}, \\
	&& \textbf{x}_2^{\text{III}} = \textbf{s}_2 + {\bf{\Phi}}\textbf{y}_{2}^{\text{I}},    
	\end{eqnarray}
\end{subequations}
where we denote the transmitter 1's transmit signals of Phase-III by $\textbf{x}_1^\text{III} \in \mathbb{C}^{M\tau_2}$ and the transmitter 2's transmit signals of Phase-III by $\textbf{x}_2^\text{III} \in \mathbb{C}^{M\tau_2}$. At the receivers, the output signals are written by
\begin{subequations}
	\begin{eqnarray}
	&& \textbf{y}_1^{\text{III}} = \textbf{H}_{2,1}^{\text{III}} \textbf{s}_2 +  \textbf{H}_{2,1}^{\text{III}} {\bf{\Phi}} \textbf{y}_2^{\text{I}} + \textbf{z}_1^\text{III},\\
	&& \textbf{y}_2^{\text{III}} = \textbf{H}_{2,2}^{\text{III}} \textbf{s}_2 +  \textbf{H}_{2,2}^{\text{III}} {\bf{\Phi}} \textbf{y}_2^{\text{I}} + \textbf{z}_2^\text{III},
	\end{eqnarray}
\end{subequations}
where we denote the receiver 1's output signals of Phase-III by $\textbf{y}_1^\text{III} \in \mathbb{C}^{N\tau_2}$, the the receiver 2's output signals of Phase-III by $\textbf{y}_2^\text{III} \in \mathbb{C}^{N\tau_2}$, and the AWGN at receivers 1 and 2 by $\textbf{z}_1^\text{III} \in \mathbb{C}^{N\tau_2}$ and $\textbf{z}_2^\text{III} \in \mathbb{C}^{N\tau_2}$, respectively. Besides, the dimension of $\textbf{H}_{i,j}^{\text{III}},i,j=1,2$ is $N\tau_2 \times M\tau_2$.  After Phase-III transmission, the receiver 1's output signals of Phase-III are fed back to the receiver 1.

\textit{Phase-IV (Interference Recurrence)}:  This phase spans $\tau_3$ TSs with the aim of simultaneous providing equations for two receivers by recurring the interference. In each TS, transmitter 1 sends output signals of Phase-II with $N$ antennas, and  transmitter 2 sends output signals of Phase-III  with $N$ antennas. As a holistic view, transmitter 1 sends a compressed  $\textbf{y}_{1}^{\text{III}}$, i.e., ${\bf{\Theta}}\textbf{y}_{1}^{\text{III}} \in \mathbb{C}^{N\tau_3}$, and transmitter 2 sends a compressed  $\textbf{y}_{2}^{\text{II}}$, i.e., ${\bf{\Theta}}\textbf{y}_{2}^{\text{II}} \in \mathbb{C}^{N\tau_3}$. Consequently, the transmitted signals of Phase-IV are expressed as
\begin{subequations}
	\begin{eqnarray}
	&& \textbf{x}_1^{\text{IV}} =   {\bf{\Theta}}\textbf{y}_{1}^{\text{III}},     \\
	&& \textbf{x}_2^{\text{IV}} =   {\bf{\Theta}}\textbf{y}_{2}^{\text{II}},    
	\end{eqnarray}
\end{subequations}
where we denote the transmitter 1's transmit signals of Phase-IV by $\textbf{x}_1^\text{IV} \in \mathbb{C}^{N\tau_3}$ and the transmitter 2's transmit signals of Phase-IV by $\textbf{x}_2^\text{IV} \in \mathbb{C}^{N\tau_3}$. At the receivers, the output signals are written by
\begin{subequations}
	\begin{eqnarray}
	&& \textbf{y}_1^{\text{IV}} =   \textbf{H}_{1,1}^\text{IV} {\bf{\Theta}}\textbf{y}_{1}^{\text{III}} + \textbf{H}_{2,1}^\text{IV} {\bf{\Theta}}\textbf{y}_{2}^{\text{II}}    + \textbf{z}_1^\text{IV},     \\
	&&\textbf{y}_2^{\text{IV}} =   \textbf{H}_{1,2}^\text{IV} {\bf{\Theta}}\textbf{y}_{1}^{\text{III}} + \textbf{H}_{2,2}^\text{IV} {\bf{\Theta}}\textbf{y}_{2}^{\text{II}}    + \textbf{z}_2^\text{IV},
	\end{eqnarray}
\end{subequations}
where we denote the receiver 1's output signals of Phase-IV by $\textbf{y}_1^\text{IV} \in \mathbb{C}^{N\tau_3}$, the the receiver 2's output signals of Phase-IV by $\textbf{y}_2^\text{IV} \in \mathbb{C}^{N\tau_3}$, and the AWGN at receivers 1 and 2 by $\textbf{z}_1^\text{IV} \in \mathbb{C}^{N\tau_3}$ and $\textbf{z}_2^\text{IV} \in \mathbb{C}^{N\tau_3}$, respectively. Besides, the dimension of $\textbf{H}_{i,j}^{\text{IV}},i,j=1,2$ is $N\tau_3 \times N \tau_3$.  

The major concern of the transmissions scheme is that we must ensure the security and decodability after transmission. To this end, we perform the security and decoding analysis in the following. Specifically, the key of security analysis is to verify the information leakage to the other receiver is zero, i.e., $I(\textbf{s}_1;\textbf{y}_2|\textbf{s}_2)=0$. However, the direct verification of zero information leakage is difficult. Hence, we resort to data processing inequality to transform it into a tractable one.

\textit{Security Analysis}: To ensure security, zero information leakage to the eavesdropper should be guaranteed. For simplicity, we define  $\textbf{y}_2 = [\textbf{y}_2^{\text{I}}; \textbf{y}_2^{\text{II}};\textbf{y}_2^{\text{III}};\textbf{y}_2^{\text{IV}}]$ and $\textbf{u} = [\textbf{u}_1;\textbf{u}_2]$.  The verification of information leakage
$I(\textbf{s}_1;\textbf{y}_2|\textbf{s}_2)=0$ is given in \eqref{Q2}, shown at the top of next page,
\begin{figure*}    
	\begin{eqnarray}
	&& I(\textbf{s}_1;\textbf{y}_2|\textbf{s}_2)  = I(\textbf{s}_1, \textbf{u};\textbf{y}_2 | \textbf{s}_2 ) - I(\textbf{u};\textbf{y}_2| \textbf{s}_1, \textbf{s}_2 ) \nonumber \\
	&& \overset{(a)}{\le} I([\textbf{H}_{1,2}^\text{I}, \textbf{H}_{2,2}^\text{I}]\textbf{u},\textbf{H}_{1,2}^\text{II}\textbf{s}_1+\textbf{H}_{1,2}^{\text{II}}{\bf{\Phi}}[\textbf{H}_{1,1}^{\text{I}}, \textbf{H}_{2,1}^{\text{I}}]\textbf{u};\textbf{y}_2 | \textbf{s}_2)  - I(\textbf{u};\textbf{y}_2| \textbf{s}_1, \textbf{s}_2) \nonumber \\
	&& \underset{\text{SNR}\rightarrow {\cal{1}}}{\overset{(b)}{=}}  \text{rk}  \left\{
	\underbrace{\begin{bmatrix}
		\textbf{I}_{N\tau_1} & \textbf{0}  \\
		\textbf{0} & \textbf{I}_{N\tau_2} \\
		\textbf{I}_{N\tau_1} & \textbf{0}  \\
		\textbf{H}_{1,2}^{\text{IV}}{\bf{\Theta}}\textbf{H}_{2,1}^{\text{III}}{\bf{\Phi}} & \textbf{H}_{2,2}^{\text{IV}}{\bf{\Theta}}
		\end{bmatrix}}_{\textbf{C}}\right\} \log \text{SNR}  \nonumber \\
	&&  - \text{rk} \left\{\underbrace{\begin{bmatrix}
		\textbf{H}_{1,2}^{\text{I}}  &  \textbf{H}_{2,2}^{\text{I}}    \nonumber \\
		\textbf{H}_{1,2}^{\text{II}}{\bf{\Phi}}\textbf{H}_{1,1}^{\text{I}}  &  \textbf{H}_{1,2}^{\text{II}}{\bf{\Phi}}\textbf{H}_{2,1}^{\text{I}}    \nonumber \\
		\textbf{H}_{2,2}^{\text{III}}{\bf{\Phi}}\textbf{H}_{1,2}^{\text{I}}  &  \textbf{H}_{2,2}^{\text{III}}{\bf{\Phi}}\textbf{H}_{2,2}^{\text{I}}    \nonumber \\
		\textbf{H}_{1,2}^{\text{IV}}{\bf{\Theta}}\textbf{H}_{2,1}^{\text{III}}{\bf{\Phi}}\textbf{H}_{1,2}^{\text{I}} + \textbf{H}_{2,2}^{\text{IV}}{\bf{\Theta}}\textbf{H}_{1,2}^{\text{II}}{\bf{\Phi}}\textbf{H}_{1,1}^{\text{I}}  &   \textbf{H}_{1,2}^{\text{IV}}{\bf{\Theta}}\textbf{H}_{2,1}^{\text{III}}{\bf{\Phi}}\textbf{H}_{2,2}^{\text{I}} + \textbf{H}_{2,2}^{\text{IV}}{\bf{\Theta}}\textbf{H}_{1,2}^{\text{II}}{\bf{\Phi}}\textbf{H}_{2,1}^{\text{I}}
		\end{bmatrix}}_{\textbf{D}}\right\} \log \text{SNR} \nonumber \\
	&& \overset{(c)}{=} N(\tau_1 + \tau_2) \log \text{SNR} - \min\{N(\tau_1+\tau_2),2N\tau_1\} \log \text{SNR}   = 0. \label{Q2}
	\end{eqnarray}
	\hrule
\end{figure*}
where the reason of each critical step is given as follows: 
\begin{enumerate}[(a)]
	\item Data processing inequality for Markov chain $(\textbf{s}_1, \textbf{u}) \rightarrow ([\textbf{H}_{1,2}^\text{I}, \textbf{H}_{2,2}^\text{I}]\textbf{u},\textbf{H}_{1,2}^\text{II}\textbf{s}_1+\textbf{H}_{1,2}^{\text{II}}{\bf{\Phi}}[\textbf{H}_{1,1}^{\text{I}}, \textbf{H}_{2,1}^{\text{I}}]\textbf{u}) \rightarrow \textbf{y}_2$.
	\item When input is circularly symmetric complex Gaussian, according to \cite{100},  rewriting into $\log\text{det}(\textbf{I} + \text{SNR}\textbf{C}\textbf{C}^H) - \log\text{det}(\textbf{I} + \text{SNR}\textbf{D}\textbf{D}^H)$, and using  Lemma 2 in \cite{20}.
	\item The rank of matrix $\textbf{C}$ is $N(\tau_1 + \tau_2)$. The rank of matrix $\textbf{D}$ is $\min\{N(\tau_1+\tau_2),2N\tau_1\}$, whose reason is given in Appendix B.
\end{enumerate}
Due to the symmetry of antenna configuration at receivers, the security analysis of the other receiver is similar. Therefore, to ensure security, according to \eqref{Q2}, shown at the top of next page, it should be followed that 
\begin{equation}
N(\tau_1 + \tau_2)  \le 2N\tau_1, \label{Z1}
\end{equation}
which is the security constraint.  Next, we perform the decoding analysis.

\textit{Decoding Analysis}: For receiver 1, the decoding equation is given as follows:
\begin{equation}
\begin{bmatrix}
\textbf{y}_1^\text{II} - \textbf{H}_{1,1}^\text{II}{\bf{\Phi}}\textbf{y}_1^\text{I} \\
\textbf{y}_1^\text{IV} - \textbf{H}_{1,1}^\text{IV}{\bf{\Theta}}\textbf{y}_1^\text{III} - \textbf{H}_{2,1}^\text{IV}{\bf{\Theta}}\textbf{H}_{1,2}^\text{II}{\bf{\Phi}} \textbf{y}_1^\text{I}
\end{bmatrix} =
\underbrace{\begin{bmatrix}
	\textbf{H}_{1,1}^\text{II} \\
	\textbf{H}_{2,1}^\text{IV}{\bf{\Theta}}\textbf{H}_{1,2}^\text{II}
	\end{bmatrix}}_{\textbf{H}_1}\textbf{s}_1 + \overline{\textbf{z}}_1, \label{H2}
\end{equation}
where $\overline{\textbf{z}}_1$ is the AWGN vector, and $\textbf{H}_1$ denotes the effective decoding channel. For receiver 2, the decoding equation is given as follows:
\begin{equation}
\begin{bmatrix}
\textbf{y}_2^\text{III} - \textbf{H}_{2,2}^\text{III}{\bf{\Phi}}\textbf{y}_2^\text{I} \\
\textbf{y}_2^\text{IV} - \textbf{H}_{2,2}^\text{IV}{\bf{\Theta}}\textbf{y}_2^\text{II} - \textbf{H}_{1,2}^\text{IV}{\bf{\Theta}}\textbf{H}_{2,1}^\text{III}{\bf{\Phi}} \textbf{y}_2^\text{I}
\end{bmatrix} =
\underbrace{\begin{bmatrix}
	\textbf{H}_{2,2}^\text{III} \\
	\textbf{H}_{1,2}^\text{IV}{\bf{\Theta}}\textbf{H}_{2,1}^\text{III}
	\end{bmatrix}}_{\textbf{H}_2} \textbf{s}_2 + \overline{\textbf{z}}_2,
\end{equation}
where $\overline{\textbf{z}}_2$ is the AWGN vector, and $\textbf{H}_2$ denotes the effective decoding channel. Accordingly, the information rate for receiver 1 is evaluated as follows:
\begin{eqnarray}
&& I(\textbf{s}_1;\textbf{y}_1|\textbf{y}_1^\text{I}) = I(\textbf{s}_1, \textbf{H}_{2,1}^\text{III}\textbf{s}_2+\textbf{H}_{2,1}^{\text{III}}{\bf{\Phi}}\textbf{y}_2^{\text{I}};\textbf{y}_1|\textbf{y}_1^\text{I}) \nonumber \\
&& - I( \textbf{H}_{2,1}^\text{III}\textbf{s}_2+\textbf{H}_{2,1}^{\text{III}}{\bf{\Phi}}\textbf{y}_2^{\text{I}};\textbf{y}_1|\textbf{s}_1,\textbf{y}_1^\text{I}) \nonumber \\
&& \underset{\text{SNR}\rightarrow {\cal{1}}}{\overset{(a)}{=}} \text{rk}\left\{\underbrace{\begin{bmatrix}
	\textbf{H}_{1,1}^\text{II} & \textbf{0} \\
	\textbf{0} & \textbf{I}_{N\tau_2} \\
	\textbf{H}_{2,1}^\text{IV}{\bf{\Theta}}\textbf{H}_{1,2}^\text{II} & \textbf{H}_{1,1}^\text{IV}{\bf{\Theta}}
	\end{bmatrix}}_{\textbf{E}}\right\} \log \text{SNR} \nonumber\\
&&  -  \text{rk}\left\{\underbrace{\begin{bmatrix}
	\textbf{I}_{N\tau_2} \\
	\textbf{H}_{1,1}^\text{\textsc{IV}}{\bf{\Theta}}
	\end{bmatrix}}_{\textbf{F}}\right\} \log \text{SNR} \nonumber \\
&& = \min\{N(\tau_1+ \tau_3 + \tau_2),M\tau_1+N\tau_2\}  \log \text{SNR} \nonumber \\
&& - N\tau_2 \log \text{SNR} \nonumber \\
&& = \min\{N(\tau_1+ \tau_3),M\tau_2\}  \log \text{SNR}, \label{J2}
\end{eqnarray}
where the reason of critical step is given as follows:
\begin{enumerate}[(a)]
	\item When input is circularly symmetric complex Gaussian, according to \cite{100},  rewriting into $\log\text{det}(\textbf{I} + \text{SNR}\textbf{E}\textbf{E}^H) - \log\text{det}(\textbf{I} + \text{SNR}\textbf{F}\textbf{F}^H)$, and using Lemma 2 in \cite{20}.
\end{enumerate}
Due to the symmetry of antenna configuration at receivers, we have $I(\textbf{s}_2;\textbf{y}_2|\textbf{y}_2^\text{I}) = \min\{N\tau_1+ N\tau_3,M\tau_2\}\log \text{SNR}$ as well. To ensure the decoding of $M\tau_2$ data symbols at each receiver, according to \eqref{J2}, it should be followed that
\begin{equation}
M\tau_2 \le N(\tau_2 + \tau_3), \label{Z2}
\end{equation}
which ensures decodability and is the decoding constraint. With the security and decoding constraints, we are able to optimize the phase duration.

\textit{Optimal Phase Duration}: We propose to optimize the phase duration by maximizing the sum-SDoF lower bound achieved by our scheme with the security and decoding constraints. In the general scheme, we securely deliver $2MN\tau_2$ data symbols over $\tau_1 + 2\tau_2 + \tau_3$ TSs. Hence, we formulate a linear-fractional optimization problem, given by
\begin{eqnarray}
\max_{\tau_1,\tau_2,\tau_3} & \dfrac{2M\tau_2}{\tau_1 + 2\tau_2 + \tau_3} \\
\text{s.t.} &  N(\tau_1 + \tau_2)  \le 2N\tau_1, \nonumber \\
&  M\tau_2 \le N(\tau_2 + \tau_3). \nonumber
\end{eqnarray}
To solve the problem, we propose to re-formulate it as follows:
\begin{eqnarray}
\max_{\tau_1/\tau_2,\tau_3/\tau_2} && \dfrac{2M}{\tau_1/\tau_2 + 2 + \tau_3/\tau_2} \\
\text{s.t.} &&  1  \le  \tau_1/\tau_2, \nonumber \\ 
&& (M-N)/N \le \tau_3/\tau_2.  \nonumber
\end{eqnarray}
Since the objective function is inversely proportional to $\tau_1/\tau_2$ and $\tau_3/\tau_2$, the optimal solution is obtained when  $\tau_1/\tau_2 = 1$ and  $\tau_3/\tau_2 = (M-N)/N$. Therefore, for the original problem, the optimal value is $2MN/(M+2N)$ by setting $\tau_1 = N$, $\tau_2 = N$, and $\tau_3 = M - N$. This shows that the following sum-SDoF lower bound is attainable: 
\begin{equation} \label{IA}
  \text{SDoF}_{\text{Sum}}^{\text{L-OF}} \ge
	\frac{2MN}{M+2N},\quad \text{if}\,\, N < M  \le 2N.
\end{equation}

\subsection{$2N < M$ Case:  The Interference Alignment Scheme with $M=2N$}

We design the transmission scheme by setting $M = 2N$ in the interference alignment transmission scheme for $N < M \le 2N$ antenna configurations, since the interference cannot be aligned if $M$ is set larger than $2N$. Therefore, for $2N<M$ antenna configurations, according to \eqref{IA}, the sum-SDoF lower bound  is given by
\begin{equation}
  \text{SDoF}_{\text{Sum}}^{\text{L-OF}} \ge  N, \quad \text{if}\,\, 2N < M.
\end{equation}
	
\section{Sum-SDoF Analysis}

\textbf{Theorem 1}: Consider the $(M,M,N,N)$  MIMO interference channel with local output feedback. The sum-SDoF lower bound is given by
\begin{equation} \label{ASDoF}
  \text{SDoF}_{\text{Sum}}^{\text{L-OF}} \ge
	\begin{cases}
		0, & M \le N/2, \\
		\dfrac{2N(2M-N)}{4M-N}, & N/2 < M \le N,\\
		\dfrac{2MN}{M+2N}, & N < M \le 2N, \\
		N, & 2N < M.
	\end{cases}
\end{equation}
\begin{IEEEproof}    
Theorem 1 is proven by the transmission schemes designed in the Section-IV. 
\end{IEEEproof}

\textbf{Corollary 1}: Consider the  $(M,M,N,N)$ MIMO interference channel with local output feedback. The sum-SDoF can be characterized in the following three cases:
\begin{equation} \label{SDoF}
  \text{SDoF}_{\text{Sum}}^{\text{L-OF}} =
	\begin{cases}
		0, & M \le N/2, \\	  
		\dfrac{2N}{3}, & N = M,   \\
		N, & 2N \le M.
	\end{cases}
\end{equation}
\begin{IEEEproof}
	According to \cite{30}, the sum-SDoF of the $(M,M,N,N)$ MIMO X channel with delayed CSIT and output feedback is given by
	\begin{equation}
		\text{SDoF}_{\text{Sum}}^{\text{XC}} = \begin{cases}
			0, & M \le N/2,\\
			\dfrac{N(2M-N)}{M}, & N/2 < M \le N, \\
			N, & N < M.
		\end{cases} \label{B1}
	\end{equation}	
	According to \cite{41}, the sum-SDoF of the $(M,M,N,N)$ MIMO interference channel with perfect CSIT is given by
	\begin{equation}
		\text{SDoF}_{\text{Sum}}^{\text{P-IC}}=  \begin{cases}
			0, & M \le N/2,\\
			4M-2N, & N/2 < M \le 2N/3, \\
			2N/3, & 2N/3 < M \le N, \\
			\dfrac{4M-2N}{3}, & N < M \le 2N, \\
			2N, & 2N < M.
		\end{cases} \label{B2}
	\end{equation}
	Since \eqref{B1} and \eqref{B2} are obtained under a better CSIT conditions than local output feedback.	Therefore, combing \eqref{B1} and \eqref{B2}, the sum-SDoF upper bound of the $(M,M,N,N)$ MIMO interference channel with local output feedback is given by
	\begin{equation} \label{U}
		\text{SDoF}_{\text{Sum}}^{\text{L-OF}} \le \begin{cases}
			0, & M \le N/2,\\
			\dfrac{N(2M-N)}{M}, & N/2 < M \le 3N/4, \\
			2N/3, & 3N/4 < M \le N, \\
			\dfrac{4M-2N}{3}, & N < M \le 5N/4, \\
			N, & 5N/4 < M,
		\end{cases}
	\end{equation}
	which comes from the existing results in \cite{29} and \cite{41}. Combining \eqref{U} and Theorem 1 completes the proof.
\end{IEEEproof}

\begin{figure}[!t]
	\centering
	\includegraphics[width=3.1in]{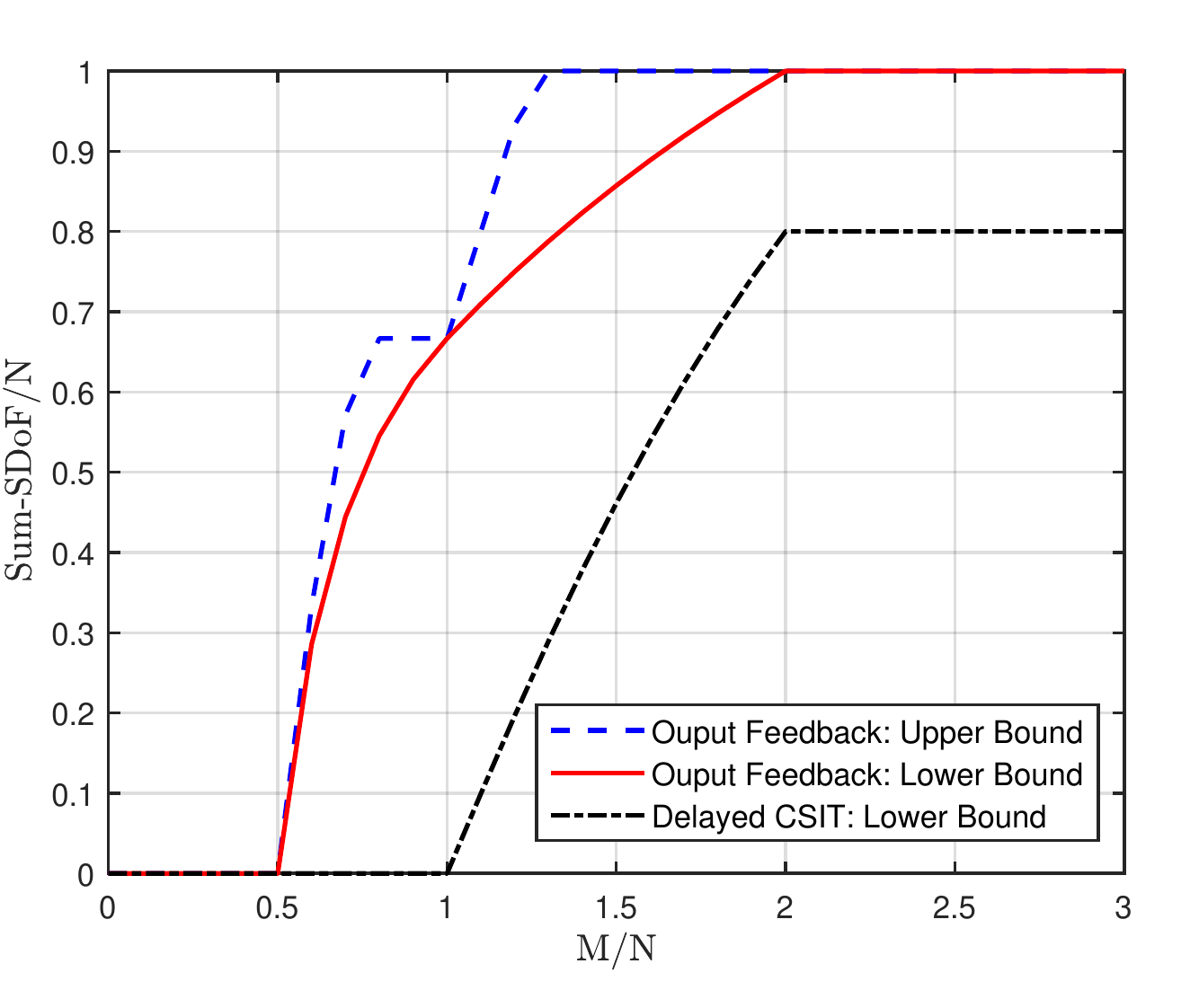}
	\caption{Derived sum-SDoF lower bound v.s. existing sum-SDoF upper bound in \eqref{U} and the sum-SDoF lower bound with delayed CSIT in \cite{29}.} \label{F2}
\end{figure}

\textit{Remark 1}: Fig. \ref{F2} shows that, the derived sum-SDoF lower bound can match the existing sum-SDoF upper bound in \eqref{U} for $M\le N/2$, $M=N$, and $2N \le M$ antenna configurations. This shows the power of output feedback. That is, using local output feedback in MIMO interference channel can achieve the same sum-SDoF as that of MIMO interference channel with perfect CSIT for $M=N$ antenna configurations and that of MIMO X channel with output feedback and delayed CSIT for $2N \le M$ antenna configurations. Furthermore, the derived sum-SDoF lower bound is higher than the state-of-the-art sum-SDoF lower bound with delayed CSIT in \cite{29}. This is because, we leverage the joint AN transmission enabled by output feedback, which doubles the ability of AN transmission, compared with the separate AN transmission in \cite{29}.

\textit{Remark 2}: Table \ref{T1} shows that, for the same $N$, enlarging $M$ will not always enlarge the sum-SDoF. Instead, the optimal $M$ occurs since $M=2N$. In other words, for a fixed $N$, the optimal $M$ to maximize the sum-SDoF is not less than $2N$. This insight suggests configuring the number of transmit antennas in the MIMO interference channel with confidential messages and local output feedback according to $M=2N$. Furthermore, Table \ref{T1} indicates that, once $M$ is smaller than $N$, the sum-SDoF will be zero.

\begin{table}[t]
	\centering
	\caption{The Value of the Sum-SDoF in \eqref{SDoF} for Some Representative Antenna Configurations} \label{T1}
	\begin{tabular}{lccccc}
		\hline 
		\rowcolor{gray!10}		 	$M \backslash N$  & $64$ &  $128$ & $256$ & $512$ & $1024$ \\ \hline 
		$64$ &  $42.6667$   & $0$     &  $0$    &   $0$  &   $0$       \\ \hline 
		\rowcolor{gray!10}		$128$&   $\textbf{64}$  &  $85.3333$   &    $0$   &   $0$   &   $0$      \\ \hline
		$256$&   $64$  &  $\textbf{128}$    &   $170.6667$     &   $0$    &   $0$       \\ \hline
		\rowcolor{gray!10}		$512$&  $64$   &   $128$   &    $\textbf{256}$   &    $341.3333$   &     $0$       \\ \hline
		$1024$&  $64$   &   $128$   &    $256$   &    $\textbf{512}$   &   $682.6667$  \\ \hline
		\rowcolor{gray!10}	 	$2048$&  $64$   &   $128$   &    $256$   &    $512$   &       $\textbf{1024}$      \\ \hline	 
	\end{tabular}
	\label{Tab3}
\end{table}

\section{Simulation Results}

In this section, we examine the
secure sum-rate performance of the transmission schemes via simulations. First,
we describe the simulation setup. Next, we present and discuss the simulation results.

 	 \begin{figure} 
	\centering
	\includegraphics[width=3in]{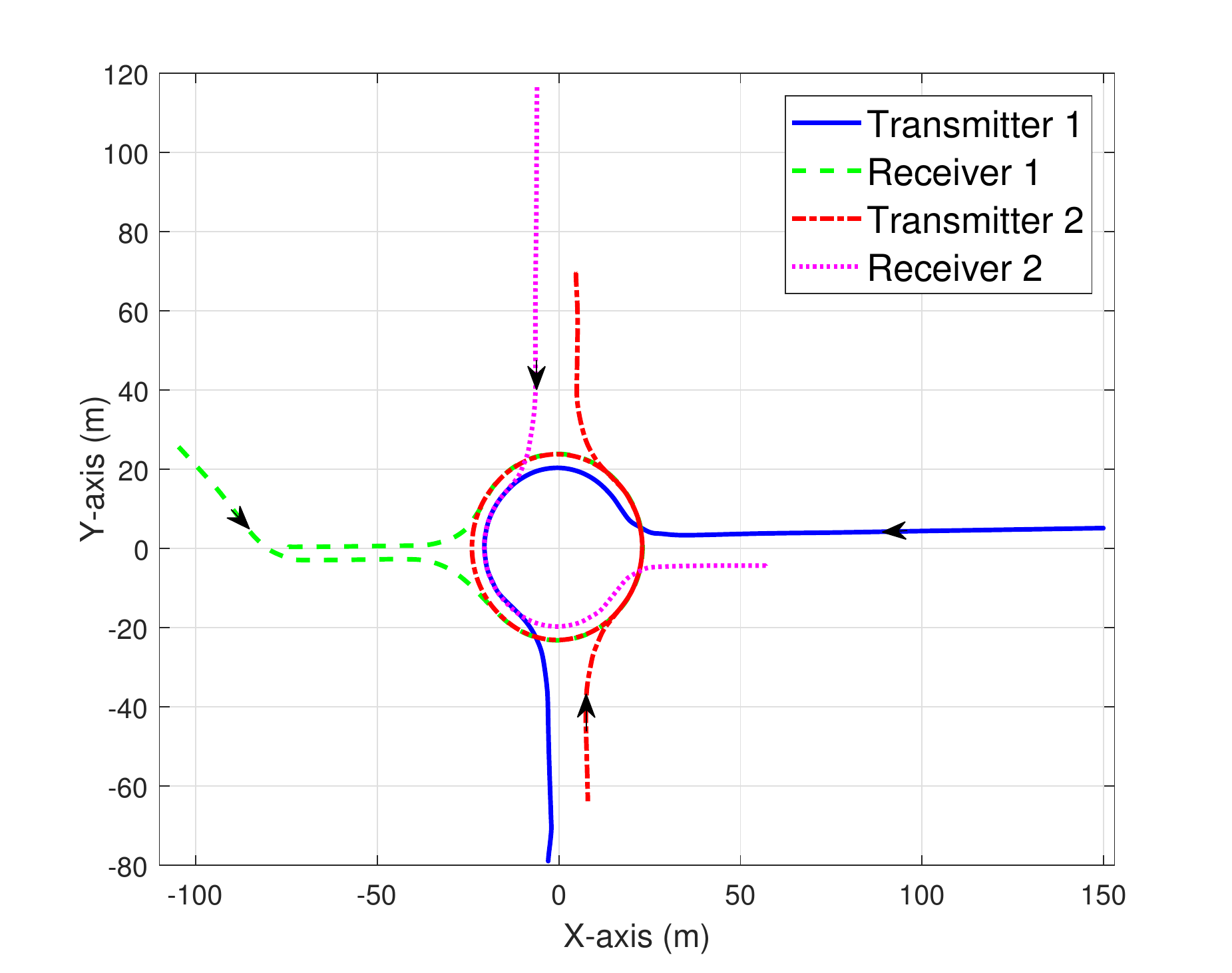}  
	\caption{Driving trajectory for 6G-enabled V2V by CARLA platform.}
	\label{SS1}
\end{figure}
\begin{figure} 
	\centering
	\includegraphics[width=2.75in]{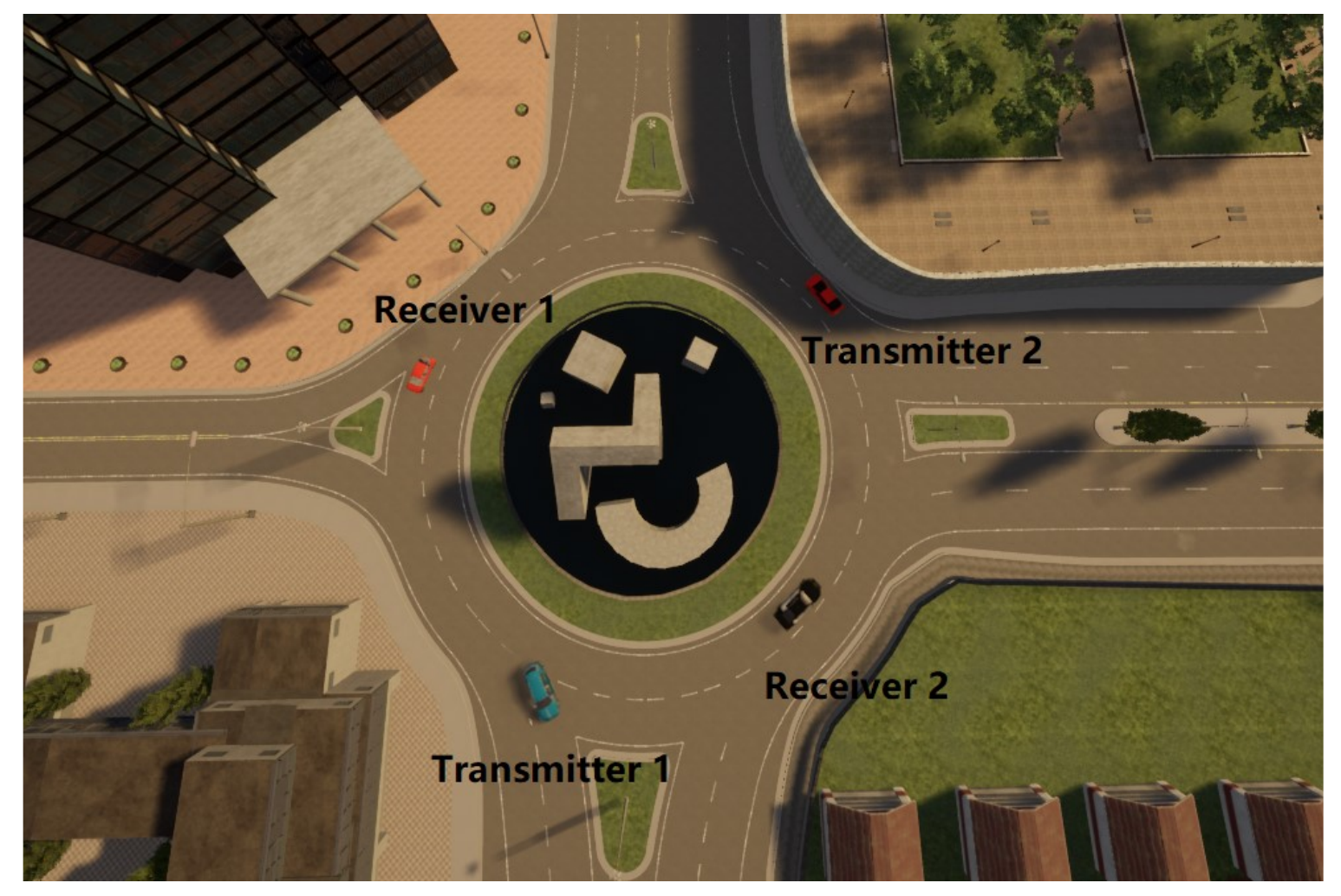}  
	\caption{Roundabout scenario from driving trajectory by CARLA platform.}
	\label{SS2}
\end{figure}

\subsection{Simulation Setup}

For the driving trajectory of 6G-enabled V2V, we resort to CARLA platform \cite{300,301,401}, which is an open-source urban driving simulator. In Fig. \ref{SS1}, we show the generated driving trajectory by CARLA, where two vehicles as transmitters and two vehicles as receivers start from four different directions and intersect at the roundabout. In Fig. \ref{SS2}, the CARLA simulation scenario at the roundabout is shown.  The large-scale fading and the small-scale fading constitute the coefficients of CSI matrices. The large-scale fading is expressed as $\eta d^{-2.5}$, where $d$ is the distance between the transmitter and the receiver, given by the generated driving trajectory, $\eta = -40$dB is the large-scale fading at $1$m distance. The small-scale fading is assumed to be Rayleigh, thus it follows zero mean unit variance complex Gaussian distribution.

 Furthermore, it is assumed that the transmit power at all the transmitters are the same and equal power allocation is employed for each transmitted data and AN symbols.  The AWGN power is $-89$dBm and equal at all the receive antennas. In the following, we adopt $10^4$ experiments to examine the secure sum-rate  performance of proposed transmission schemes, in order to average out the fluctuation of small-scale fading.

\subsection{Results and Discussion}

\textit{IA-D as a Comparison Scheme}: The interference alignment with delayed CSIT (i.e., IA-D), given in \cite{29}, is described as follows:  The IA-D utilizes delayed CSIT and has 5 phases. In Phase-I, the transmitter 1 sends AN. In Phase-II, the transmitter 2 sends AN. The function of remaining three phases is the same as that in our interference alignment scheme. Hence, its effective decoding channel is also the same as ours.

%\textit{Evaluation of Symbol Error Rate}: The symbol error rate of proposed transmission schemes is provided in Fig. \ref{SS3} and Fig. \ref{SS4} via Monte-Carlo simulations, where minimum mean square error (MMSE) or zero-forcing (ZF) detector, and 4-quadrature amplitude modulation (QAM) are employed.} 

%Fig. \ref{SS3} shows that a marginal variance of symbol error
%rate as $M$ and $N$ dramatically increase, when the interference decoding scheme is adopted. This is because, the interference decoding scheme is designed for multiplexing only, and the diversity-oriented design is not considered. This suggests to embed the diversity-oriented design, such as space-time coding, to our scheme to enhance the error rate performance, e.g., \cite{260} does. Moreover, the IA-D is infeasible due to $M=N$.} 

%Fig. \ref{SS4} shows that the interference alignment scheme and the IA-D have the same symbol error rate, when $M$ and $N$ dramatically increase. This is because, the effective decoding channel of our scheme and IA-D are the same. Likewise, since only multiplexing is considered, symbol error rate decreases marginally, when $M$ and $N$ dramatically increase. } 

\textit{Evaluation of the Secure Sum-rate}:  The secure sum-rate of proposed transmission schemes is provided in Fig. \ref{SS5} and Fig. \ref{SS6} via Monte Carlo simulations. The desired information rate minuses leakage information rate is the secure rate. The secure sum-rate is the sum of secure rate at each receiver.

Fig. \ref{SS5} shows that the secure sum-rate increases with $M$ and $N$. Especially, for high SNR regime, the difference of secure sum-rate for different $M$ and $N$ is larger. This is because, the larger $M$ and $N$ imply a higher sum-SDoF lower bound, which approximates the secure sum-rate in high SNR regime. This suggests equipping with more antennas at the transmitters and receivers to increase the secure sum-rate, especially when the transmit power is high.

Fig. \ref{SS6} shows that the secure sum-rate of our interference alignment scheme is higher than that of IA-D, for the same $M$ and $N$. This is because, our interference alignment scheme adopts the joint AN transmission, while the IA-D employs the separate AN transmission. Therefore, we save one phase for AN transmission, which leads to a  higher secure sum-rate. This suggests employing our interference alignment scheme in the presence of delayed feedback.

\section{Conclusions}

For the $(M,M,N,N)$ MIMO interference channel with confidential messages and local output feedback, we have proposed two novel transmission schemes, i.e., the interference decoding scheme and the interference alignment scheme. We have analyzed the security and decoding constraints of the proposed schemes. The linear-fractional optimization problem has been established to maximize the sum-SDoF lower bound achieved by the schemes. The simulation results have validated that the secure sum-rate of our scheme is superior to  that of state-of-the-art scheme with delayed CSIT. Theoretically, the main implications are summarized as follows: A  non-trivial sum-SDoF lower bound for all antenna configurations has been derived, and is the sum-SDoF for $N/2 \le M$, $N=M$, and $2N \le M$ antenna configurations. Moreover, for a fixed $N$, the optimal $M$ to maximize the sum-SDoF is not less than $2N$. Practically, the main implication is that using local output feedback can lead to a higher secure sum-rate than that by using delayed CSIT. Following this work, there can be several future directions, listed as follows:

\begin{itemize}
	\item One extension of this work is to study the sum-SDoF of the $(M,M,N,N)$ MIMO interference channel with local output feedback and an external eavesdropper or a cooperative jammer. Another extension is to investigate the sum-SDoF of the $(M,M,N,N)$ MIMO interference channel with local output feedback and common message.
	
	\begin{figure}[t]
		\centering
		\includegraphics[width=3in]{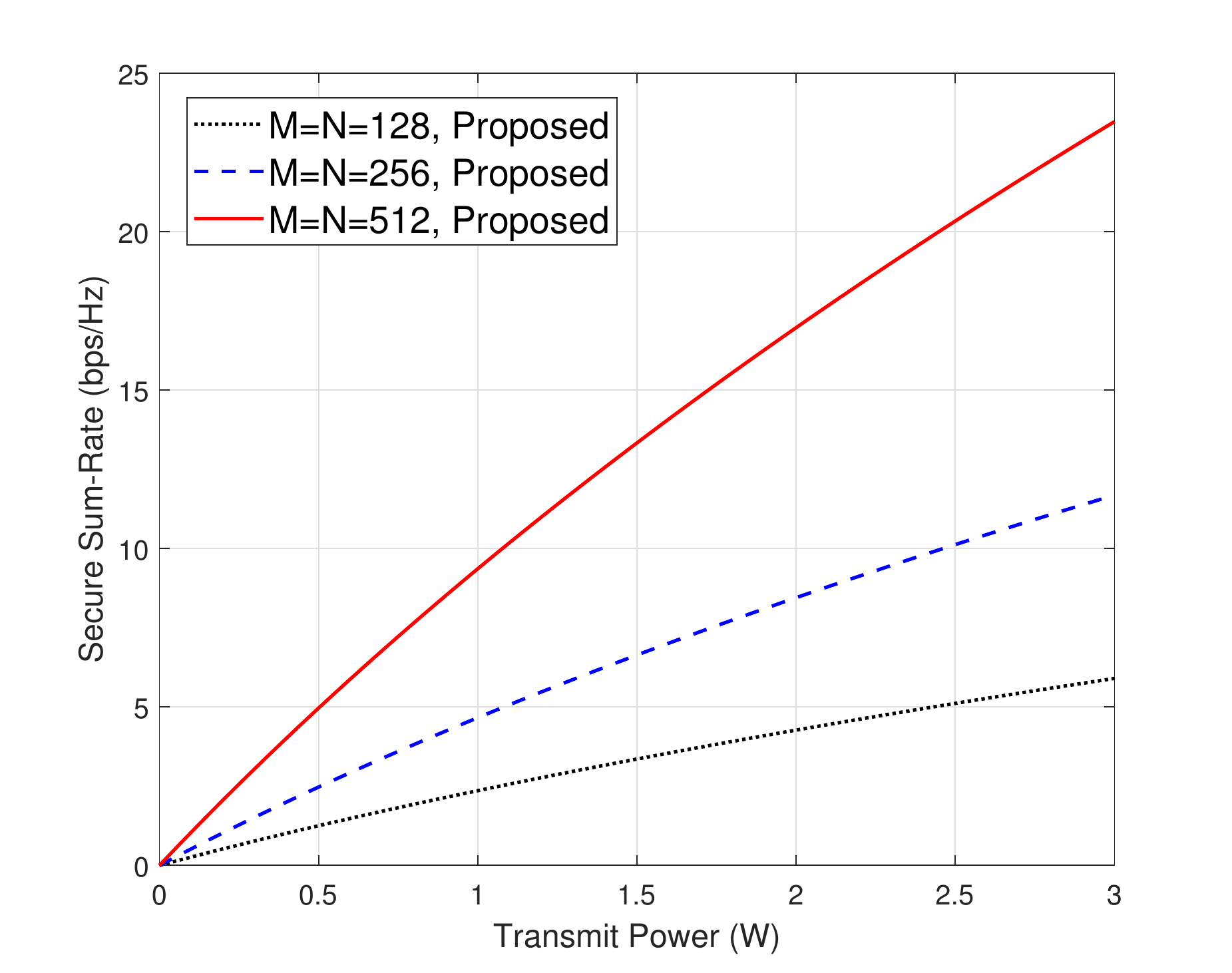}
		\caption{Secure sum-rate of the interference decoding scheme.} \label{SS5}
	\end{figure} 
		\begin{figure}[t] 
	\centering
	\includegraphics[width=3in]{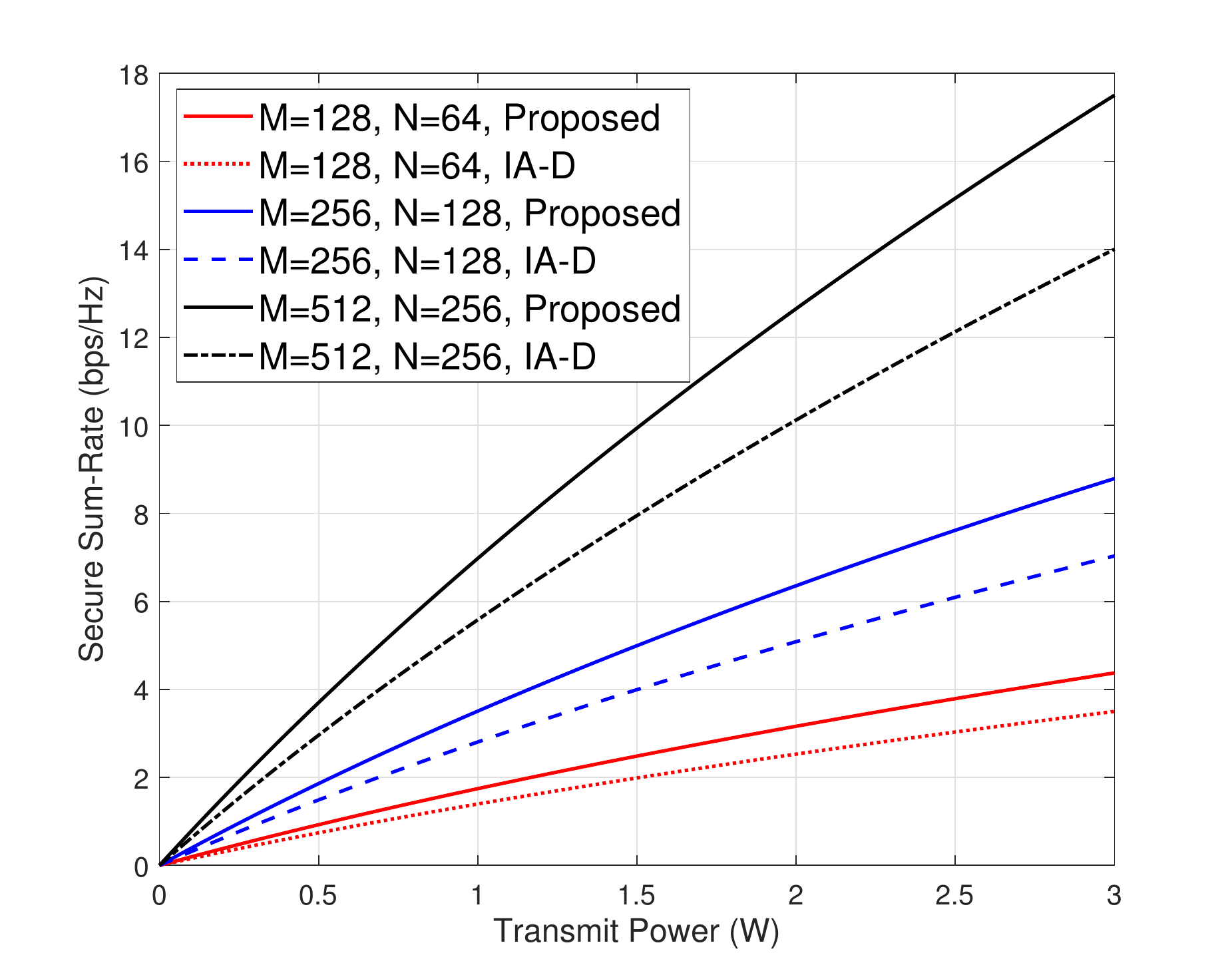}
	\caption{Secure sum-rate: Interference alignment scheme v.s. IA-D.} \label{SS6}
\end{figure}
	
	\item The tight sum-SDoF upper bound of the $(M,M,N,N)$ MIMO interference channel with local output feedback can be studied. The major difficulties lie in how to design genie signalings that can both facilitate the upper bound derivation and do not increase the sum-SDoF, and whether the statistical equivalence property (widely being used to establish the upper bound, see \cite{21,22,23,24}) holds for the $(M,M,N,N)$ MIMO interference channel with confidential messages and local output feedback or not.  
	\item  The sum-SDoF of the   $(M_1,M_2,N_1,N_2)$ MIMO interference channel with local output feedback can be investigated, where the transmitter $i=1,2$ has $M_i$ antennas and the receiver $j=1,2$ has $N_j$ antennas.  In addition to the challenge of tight upper bound establishment, the major difficulty lies in designing and analyzing the transmission schemes for numerous antenna configuration cases (see \cite{01,02} for the DoF without security), due to the lack of symmetry. For some asymmetry antenna configurations, the transmission design and analysis is different from that in symmetry antenna configurations. 
	\item The sum-SDoF of the $K$-user MIMO interference channel with  local output feedback can be investigated. Due to the multi-sources of interference (see \cite{250,251,252} for the DoF without security), the method is largely different from that in two-user case, i.e., MIMO interference channel. Hence, new techniques are needed.
\end{itemize}

%		\begin{figure}[!t]
%	\centering
%	\includegraphics[width=3in]{S1_SDoF}
%	\caption{Symbol error rate of the interference decoding scheme.}} \label{SS3}
%\end{figure} 
%\begin{figure}[!t] 
%	\centering
%	\includegraphics[width=3in]{S2_SDoF}
%	\caption{Symbol error rate: Interference alignment scheme v.s. IA-D.}} \label{SS4}
%\end{figure} 

\section{Acknowledgments}

T. Zhang would like to thank his PhD supervisor Prof. P. C. Ching for his guidance during past 4.5 years at CUHK.

	\begin{appendices}
		\section{Rank Analysis in $N/2 < M \le N$ Case}

		The rank analysis of matrix $\textbf{B}$ is given as follows: Note that we can represent matrix $\textbf{B}$ by multiplication of the following two matrices:
		\begin{equation}
		\underbrace{\begin{bmatrix}
			\textbf{I}_{N\tau_1} & \textbf{0} \\
			\textbf{H}_{2,2}^{\text{II}}{\bf{\Phi}}_2^{a} & \textbf{H}_{1,2}^{\text{II}}{\bf{\Phi}}_1^{a} \\
			\textbf{H}_{2,2}^{\text{III}}{\bf{\Phi}}_2^{b}  & \textbf{H}_{1,2}^{\text{III}}{\bf{\Phi}}_1^{b}
			\end{bmatrix}}_{\textbf{U}}
		\underbrace{\begin{bmatrix}
			\textbf{H}_{1,2}^{\text{I}}  & \textbf{H}_{2,2}^{\text{I}} \\
			\textbf{H}_{1,1}^{\text{I}}  & \textbf{H}_{2,1}^{\text{I}}
			\end{bmatrix}}_{\textbf{L}}. \nonumber
		\end{equation}
		Using Gaussian elimination,  we re-write matrix $\textbf{U}$  as
		\begin{equation}
		\underbrace{\begin{bmatrix}
			\textbf{I}_{N\tau_1} & \textbf{0} \\
			\textbf{0} & \textbf{H}_{1,2}^{\text{II}}{\bf{\Phi}}_1^{a} \\
			\textbf{0}  & \textbf{H}_{1,2}^{\text{III}}{\bf{\Phi}}_1^{b}
			\end{bmatrix}}_{\widetilde{\textbf{U}}}. \nonumber
		\end{equation}
		Clearly, to determine the rank of matrix $\widetilde{\textbf{U}}$, we should decide the rank of matrix $[\textbf{H}_{1,2}^{\text{II}}{\bf{\Phi}}_1^{a};\textbf{H}_{1,2}^{\text{III}}{\bf{\Phi}}_1^{b}]$ first, which can be decomposed into
		\begin{equation}
		\begin{bmatrix}
		\textbf{H}_{1,2}^{\text{II}}{\bf{\Phi}}_1^{a} \\
		\textbf{H}_{1,2}^{\text{III}}{\bf{\Phi}}_1^{b}
		\end{bmatrix} = \underbrace{\begin{bmatrix}
			\textbf{H}_{1,2}^{\text{II}} & \textbf{0} \\
			\textbf{0} & \textbf{H}_{1,2}^{\text{III}}
			\end{bmatrix}}_{\textbf{P}}
		\underbrace{\begin{bmatrix}
			{\bf{\Phi}}_1^{a} \\
			{\bf{\Phi}}_1^{b}
			\end{bmatrix}}_{\textbf{Q}}.
		\end{equation}
		The rank of matrix $\textbf{P}$ is $N\tau_2$ and the rank of matrix $\textbf{Q}$ is
		$\min\{N\tau_2,N\tau_1\}$. Since the rank of two matrix multiplication is determined by the minimal rank of individual matrix \cite{65}, thus the rank of matrix $[\textbf{H}_{1,2}^{\text{II}}{\bf{\Phi}}_1^{a};\textbf{H}_{1,2}^{\text{III}}{\bf{\Phi}}_1^{b}]$ is $\min\{N\tau_2, N\tau_1\}$. Therefore, the rank of matrix $\widetilde{\textbf{U}}$ is $\min\{N(\tau_1 + \tau_2), 2N\tau_1\}$. Since matrix $\textbf{U}$ has the same rank as matrix $\widetilde{\textbf{U}}$, the rank of matrix $\textbf{U}$ is also $\min\{N(\tau_1 + \tau_2), 2N\tau_1\}$.  It can be easily see that the rank of matrix $\textbf{L}$ is $2M\tau_1$. Due to $M \le N$, the rank of  matrix $\textbf{B}$ is $\min\{N(\tau_1 + \tau_2),2M\tau_1\}$.

		\section{Rank Analysis in $N < M \le 2N$ Case}
		The rank analysis of matrix $\textbf{D}$ is given as follows: Note that we can represent matrix $\textbf{D}$ by multiplication of the following two matrices:
		\begin{equation}
		\underbrace{\begin{bmatrix}
			\textbf{I}_{N\tau_1} & \textbf{0} \\
			\textbf{0} & \textbf{H}_{1,2}^{\text{II}}{\bf{\Phi}} \\
			\textbf{H}_{2,2}^{\text{III}}{\bf{\Phi}} & \textbf{0} \\
			\textbf{H}_{1,2}^{\text{IV}}{\bf{\Theta}}\textbf{H}_{2,1}^{\text{III}}{\bf{\Phi}}  & \textbf{H}_{2,2}^{\text{IV}}{\bf{\Theta}}\textbf{H}_{1,2}^{\text{II}}{\bf{\Phi}}
			\end{bmatrix}}_{\textbf{U}}
		\underbrace{\begin{bmatrix}
			\textbf{H}_{1,2}^{\text{I}}  & \textbf{H}_{2,2}^{\text{I}} \\
			\textbf{H}_{1,1}^{\text{I}}  & \textbf{H}_{2,1}^{\text{I}}
			\end{bmatrix}}_{\textbf{L}}. \nonumber
		\end{equation}
		Since  the rank of two matrix multiplication is determined by the minimal rank of individual matrix \cite{65}, we next decide the rank of individual matrix. To determine the rank of matrix $\textbf{U}$, we should decide the
		rank of matrix $\textbf{H}_{1,2}^{\text{II}}{\bf{\Phi}}$ first, whose rank is determined by $\textbf{H}_{1,2}^{\text{II}}$ and ${\bf{\Phi}}$. The rank of matrix $\textbf{H}_{1,2}^{\text{II}}$ is $N\tau_2$ and the rank of matrix ${\bf{\Phi}}$ is $\min\{M\tau_2,N\tau_1\}$. Consequently, the rank of matrix $\textbf{H}_{1,2}^{\text{II}}{\bf{\Phi}}$ is $\min\{N\tau_1,N\tau_2\}$, which implies that the rank of  matrix $\textbf{U}$ is $\min\{N(\tau_1+\tau_2),2N\tau_1\}$. It can be easily seen that the rank of matrix $\textbf{L}$ is $2N\tau_1$.  Therefore, the rank of matrix $\textbf{D}$ is $\min\{N(\tau_1+\tau_2),2N\tau_1\}$.

	\end{appendices}

	\bibliographystyle{IEEEtran}
	\bibliography{IC_SDoF}

\end{document}